\documentclass[journal]{IEEEtran}

\usepackage{cite}
\usepackage{hyperref}

\ifCLASSINFOpdf
    \usepackage[pdftex]{graphicx}
    \graphicspath{{./figures/}}
\else
\fi

\usepackage{mathtools}
\usepackage{algorithm}
\usepackage{algorithmic}
\usepackage{eqparbox}

\usepackage{xcolor}
\usepackage{xspace} 

\ifCLASSOPTIONcompsoc
    \usepackage[caption=false,font=normalsize,labelfont=sf,textfont=sf]{subfig}
\else
  \usepackage[caption=false,font=footnotesize]{subfig}
\fi
\usepackage{siunitx}
\usepackage{mathptmx}
\usepackage{threeparttable}
\usepackage{amsmath,amssymb,amsfonts}
\usepackage{algorithmic}
\usepackage{textcomp}
\usepackage{array}
\usepackage{multirow}

\begin{document}
\bstctlcite{IEEEexample:BSTcontrol} 

\title{Omnidirectional Wireless Power Transfer for Millimetric Magnetoelectric Biomedical Implants}

\author{
        Wei Wang,~\IEEEmembership{Student Member,~IEEE,}
        Zhanghao~Yu,~\IEEEmembership{Student Member,~IEEE,}
        Yiwei~Zou,~\IEEEmembership{Student Member,~IEEE,}
        Joshua~E.~Woods,~\IEEEmembership{Student Member,~IEEE,}
        Prahalad~Chari, 
        Yumin~Su,~\IEEEmembership{Student Member,~IEEE,}
        Jacob~T.~Robinson,~\IEEEmembership{Senior~Member,~IEEE,}
        and~Kaiyuan~Yang,~\IEEEmembership{Member,~IEEE}
\thanks{Manuscript received on}
\thanks{This work was supported in part by the National Science Foundation (NSF) CAREER program under award 2146476, the Robert and Janice McNair Foundation, McNair Medical Institute, the National Institutes of Health NIH under grant U18EB029353, and the Defense Advanced Research Projects Agency (DARPA) under Agreement FA8650-21-2-7119. (Corresponding author: Kaiyuan Yang, kyang@rice.edu)}
\thanks{W. Wang, Z. Yu, Y. Zou, J. E. Woods, P. Chari, Y. Su, and K. Yang are with the Department of Electrical and Computer Engineering, Rice University, Houston TX, 77005, USA.}
\thanks{J. T. Robinson is with the Department of Electrical and Computer Engineering, Rice University, Houston TX, 77005, USA, and Baylor College of Medicine, Houston, TX 77030, USA.}
\thanks{J. E. Woods, J. T. Robinson, and K. Yang receive monetary and/or equity compensation from Motif Neurotech.}}

\markboth{Journal of Solid-State Circuits}%
{Shell \MakeLowercase{\textit{et al.}}: A Sample Article Using IEEEtran.cls for IEEE Journals}


\maketitle

\begin{abstract}
Miniature bioelectronic implants promise revolutionary therapies for cardiovascular and neurological disorders. 
Wireless power transfer (WPT) is a significant method for miniaturization, eliminating the need for bulky batteries in today's devices. Despite successful demonstrations of millimetric battery-free implants in animal models, the robustness and efficiency of WPT are known to degrade significantly under misalignment incurred by body movements, respiration, heart beating, and limited control of implant orientation during surgery. 
This paper presents an omnidirectional WPT platform for millimetric bioelectronic implants, employing the emerging magnetoelectric (ME) WPT modality, and ``magnetic field steering" technique based on multiple transmitter (TX) coils. To accurately sense the weak coupling in a miniature implant and adaptively control the multi-coil TX array in a closed loop, we develop an Active Echo (AE) scheme using a tiny coil on the implant. 
Our prototype comprises a fully integrated 14.2mm$^3$ implantable stimulator embedding a custom low-power System-on-Chip (SoC) powered by an ME film, a transmitter with a custom three-channel AE RX chip, and a multi-coil TX array with mutual inductance cancellation. The AE RX achieves -161dBm/Hz input-referred noise with 64dB gain tuning range to reliably sense the AE signal, and offers fast polarity detection for driver control. 
%
AE simultaneously enhances the robustness, efficiency, and charging range of ME WPT. Under 90\textdegree{} rotation from the ideal position, our omnidirectional WPT system achieves 6.8$\times$ higher power transfer efficiency (PTE) than a single-coil baseline. The tracking error of AE negligibly degrades the PTE by less than 2\% from using ideal control.
\end{abstract}


\section{Introduction}

\IEEEPARstart{B}{ioelectronic} implants have played a vital role in helping millions of patients with cardiovascular and neurological disorders~\cite{verrills_review_2016,rush_vagus_nodate,kim_advances_2023} since the first pacemaker developed in the 1950s. With recent advances in bio-engineering, neuroscience, and medical research, bioelectronic implants are expected to further transform healthcare and medicine. Decades of efforts have been dedicated to miniaturizing these bio-implants~\cite{nair_miniature_2023}, which reduces infection risks, simplifies surgical procedures, improves patient acceptance, and opens up new applications~\cite{najafiaghdam_3d_2022}. 
Today, batteries still dominate the volume of commercially available bio-implants. Replacing the batteries with wireless power transfer (WPT) technologies promises a significant leap forward in miniaturizing the implants to the millimeter cubic regime.
Fully integrated millimetric battery-free implantable devices have been demonstrated with various WPT modalities, e.g., inductive coupling~\cite{jia_mm-sized_2019, habibagahi_vagus_2022,park_wireless_2021,lee_neural_2021, burton_wireless_2020}, RF~\cite{leung_distributed_2019}, ultrasound~\cite{johnson_stimdust_2018,charthad_mm-sized_2018, muller_0013_2012, rabbani_173_2024}, light~\cite{lim_light_2021} and magnetoelectric (ME)~\cite{singer_magnetoelectric_2020, Yu_343_ISSCC, chen_wireless_2022, Mehdi_Comperhensive_2021, Mehdi_MagSonic_2024}. 
{State-of-the-art millimetric bio-implants have been achieved for various applications requiring an implantation depth of no more than 5-6 centimeters, including cortical~\cite{Josh_human}, spinal cord~\cite{chen_wireless_2022}, vagus nerve~\cite{habibagahi_vagus_2022}, peripheral nerve~\cite{johnson_stimdust_2018}, neural stimulators, and leadless pacemakers~\cite{yu_magnetoelectric_2022, lyu2020synchronized}. For applications that require more than 10 centimeters of power transfer distance within the body, such as ingestible devices~\cite{ramadi_bioinspired_2023, sharma_location-aware_2023}, existing WPT technologies cannot replace batteries yet, mainly because of safety limits on the maximum transmitter power. }

However, the power transfer efficiency (PTE) of WPT is known to severely degrade by misalignment between an external transmitter and a miniaturized receiver, which is almost inevitable during chronic implantation, due to body movements, respiration, heart beating, and limited control of implant orientation during surgeries.
%
%

\begin{figure}[t!]
\centering
\includegraphics[scale = 0.63]{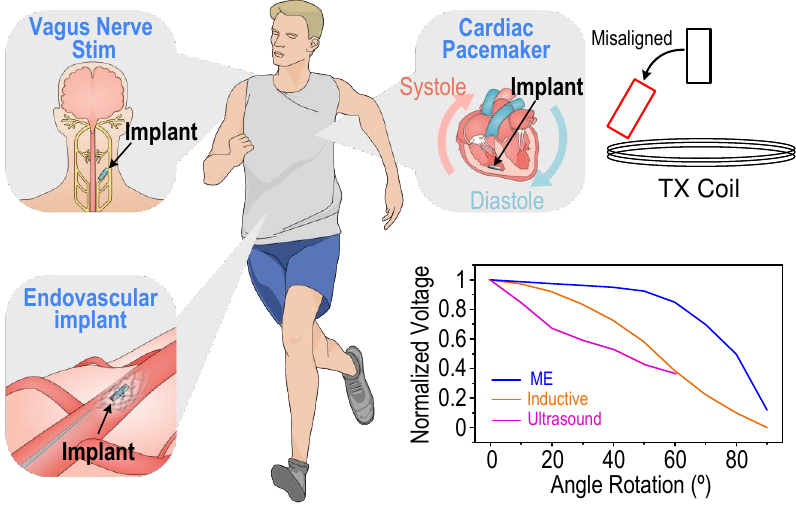}
\vskip -2ex
\caption{Selected applications of wirelessly powered bio-implants and the sensitivity to angular rotations of various WPT modalities \cite{burton_wireless_2020,piech_wireless_2020,yu_magnetoelectric_2022}.}
\label{Fig_1}
\end{figure}

Fig.~\ref{Fig_1} showcases the angular sensitivity of three popular WPT modalities: inductive coupling, ultrasound, and magnetoelectric. Inductive coupling is the most well-studied wireless charging method~\cite{burton_wireless_2020}. However, it faces two major challenges for miniature implants, including the PTE's cubical reduction with the receiver's radius~\cite{yu_magni_2020} and the increased specific absorption rates (SAR) in tissues when adopting a higher frequency for coil size reduction~\cite{noauthor_iecieee_nodate}. 
Ultrasonic WPT circumvents the body absorption issue with mm-scale transducers acoustically resonant at MHz, delivering higher power with superior efficiency to miniature implants~\cite{piech_wireless_2020}. 
However, it suffers from reflections through different mediums and exhibits high sensitivity to rotations~\cite{piech_wireless_2020} and lateral mismatch with focused ultrasound TX~\cite{Gourdouparis_62_2024_focus_US}.
Lastly, the emerging ME WPT transducers convert an AC magnetic field to electrical voltage via acoustic coupling, which enjoys higher PTE than inductive coupling with mm-scale receivers while avoiding the penetration issues in ultrasound~\cite{yu_magni_2020}. 
More importantly, thanks to the magnetic flux concentration effects~\cite{singer_wireless_2021,yu_magnetoelectric_2022}, ME naturally offers superior misalignment tolerance. 

Adaptation to the actual position of the implant is a clear pathway to address the misalignment tolerance. 
Closed-loop power regulation has been applied to inductive~\cite{yan_lu_1356mhz_2013,tang_336_2021} and ME WPT systems~\cite{yu_wireless_2022}, where the implant reports its received power through a data uplink back to the external TX for adjusting output power. 
However, this approach cannot handle severe lateral and angular offsets and burns significantly higher power of the transmitter at larger misalignment. 


\begin{figure}[t!]
\centering
{\includegraphics[scale = 0.9]{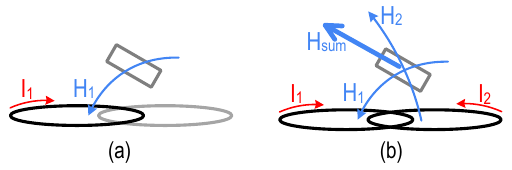}%
}
\vskip -2ex
\caption{
{Magnetic field distribution at the implant's site when (a) turning on one TX coil, and (b) optimally configuring the two coils.} }
\label{fig_magnetic_beam}
\end{figure}


\begin{figure}[t!]
\centering
\includegraphics[width = \linewidth]{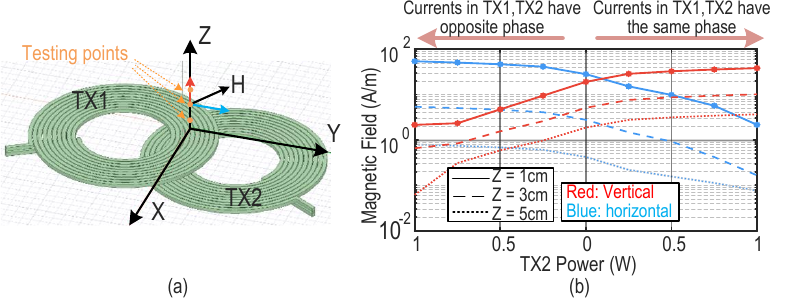}%
\vskip -2ex
\caption{
{HFSS simulation setup and results of magnetic field steering: the power of TX1 is 1W, and the power of TX2 changes from 1W with the opposite current phase to 1W with the same phase.} }
\vskip -3ex
\label{field_steering}
\end{figure}

\begin{figure}[t!]
\centering
\includegraphics[scale = 0.65]{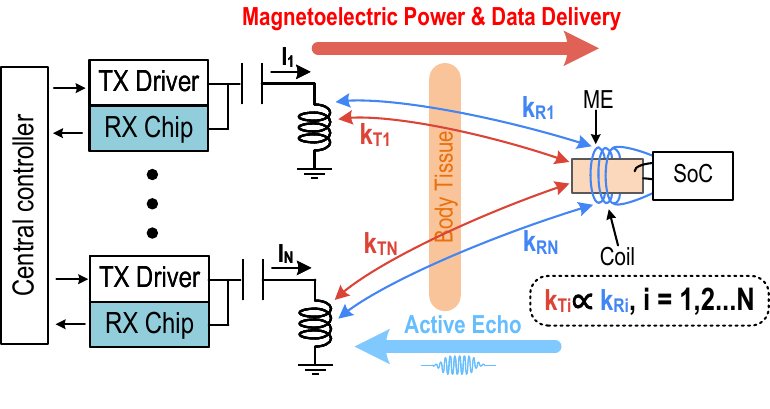}
\vskip -2ex
\caption{Principle of Active Echo (AE) to enable adaptive omnidirectional magnetoelectric wireless power transfer.}
\label{fig_2}
\vskip -3ex
\end{figure}

In order to fundamentally alleviate the PTE reduction and robustness issues in magnetic WPT systems (both inductive and ME), an intuitive solution is to adapt the magnetic field generated by the TX to better align with the receiver. 
As shown in Fig.~\ref{fig_magnetic_beam}~a, a single TX coil does not possess the degree of freedom to adapt to the RX position, but it is possible with multiple coils. As an example, the two coils in Fig.~\ref{fig_magnetic_beam}~b could manipulate the direction and intensity of the induced magnetic field at a given position by adjusting the phase and amplitude of each coil's current. This is known as ``magnetic field steering". 
{Fig.~\ref{field_steering} proves that the magnetic field can operate with a minimum distance of 5cm when the diameter of TX coils is 4.2cm, which is adequate for our intended applications.}
By adding more TX coils, the degree of freedom for steering is increased, leading to higher accuracy and broader coverage. This concept has been demonstrated on inductive WPT systems \cite{jadidian_magnetic_2014,shi_wireless_2015,qiu_678-mhz_2022,huang_308_2023}, with various techniques for controlling the multi-coil array. 
%
%
However, all existing designs were developed for applications with larger RX coils (i.e., tens of centimeters in diameter) strongly coupled to TX coils of similar sizes, such as phone charging. 
They cannot be adopted for a millimeter-scale receiver since the coupling is too weak to sense on the TX coil. Blindly searching for the optimal setting is possible with brute force or gradient algorithms, like blind adaptive beamforming~\cite{chen_multiantenna_2016}, but will incur excessive latency, energy loss, and interruptions to normal operation. It also scales badly with the number of TX coils. 

\begin{figure}[t!]
\centering
\includegraphics[width=\linewidth]{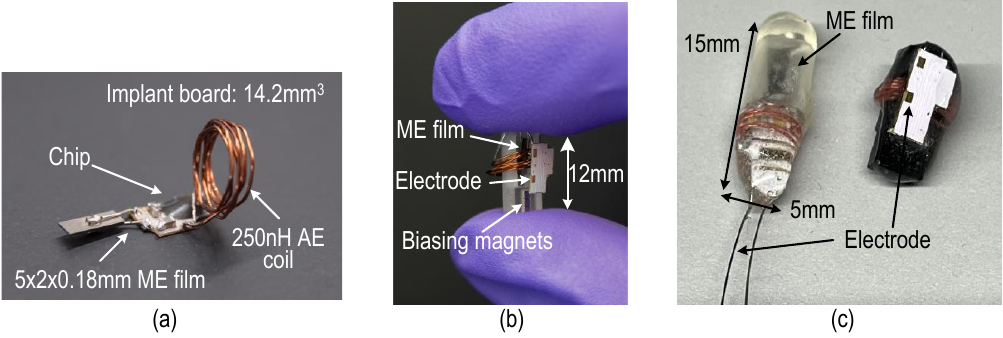}
\vskip -1ex
\caption{Prototypes of implant: (a) Unfolded implant board with storage cap, ME film, and AE coil assembled; (b) Implant with an mm-scale plastic package and aligned ME film and AE coil; (3) Implant encapsulated with medical epoxy for cardiac pacing and neurostimulation.}
\label{fig_3}
\end{figure}

To tackle this challenge, we present the Active Echo (AE) coupling sensing technique, which enables an omnidirectional WPT platform for millimetric implants~\cite{wang_171_2024}, as depicted in Fig.~\ref{fig_2}. By actively transmitting an AE tone with a tiny auxiliary coil on the implant, we are able to indirectly sense the effective coupling between the external TX coil and the implant's power receiver (an ME film in our system) and control multi-coil magnetic field steering in a real-time closed loop. 
Our prototype system comprises 1) a 14.2mm$^3$ implantable bio-stimulator, assembled with a 5$\times$2 $mm^2$ ME film for WPT, a 250nH AE coil, and a fully integrated low-power SoC, as shown in Fig.~\ref{fig_3}; 2) a complete external transmitter with GaN drivers and a multi-channel AE RX chip featuring -161dBm/Hz input-referred noise for sensing coupling coefficients; 3) a mutual inductance canceled TX coil array, achieving less than -50dB S21 between each pair of coils to minimize resonant frequency shifts and resonant current reduction. The overall system achieves a 6.8$\times$ PTE improvement over a single-coil baseline at a 90\textdegree{} rotation from the ideal alignment, while having a less than 2\% efficiency loss due to the AE tracking error.

The rest of this article is organized as follows: Section II illustrates the principles of the proposed omnidirectional WPT system and AE technique. Section III provides implementation details of the circuit and system designs. Section IV presents the experimental results, including chip measurement and system in-vitro tests. Section V concludes this article.

\section{Principles of Omnidirectional WPT}
\subsection{Multi-Coil Adaptive Power Transfer via Active Echo}

Similar to inductive coupling, the received power by an ME transducer is proportional to the magnetic field along the long axis of the rectangular-shaped thin-film composite. 
To optimize PTE with adaptive magnetic field steering, the straightforward approach is to find the receiver's pose in real time and use it to calculate the optimal TX coil current allocation that best matches the desired field condition, using electromagnetic theories. Localizing implants is feasible with triangulation methods~\cite{Ruston_wireless_3D_2022, Sharma_3D_surgical_2020}.
However, they can only measure the positions in 3 degrees of freedom (DoF), assuming no rotations, and demand extra computation power and time. Employing two or three orthogonal sensors on the implant can achieve six DoF~\cite{huang_im6d_2015,sharma_location-aware_2023}, but this makes the implant bulky and consumes more power and area. The process of estimating poses for the implant is not lightweight due to the uplink and synchronization of multiple sets of sensor data. The stability and SNR of the uplink data also affect the success rate of pose estimation, leading to additional communication requirements of the implant.

\begin{figure}[t!]
\centering
\includegraphics[width=0.72\linewidth]{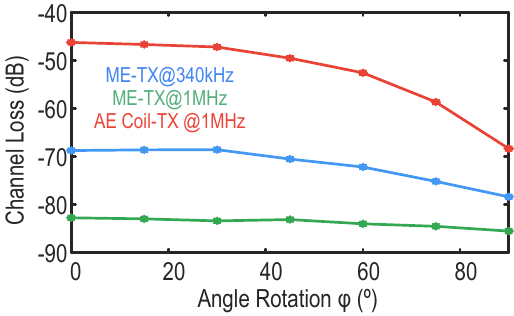}
\vskip -1ex
\caption{
{Reverse channel loss when ME and AE coil are placed 20mm above the center of the TX coil.} }
\label{AE_ME_revise}
\end{figure}

\begin{figure}[t!]
\centering
\includegraphics[scale = 0.72]{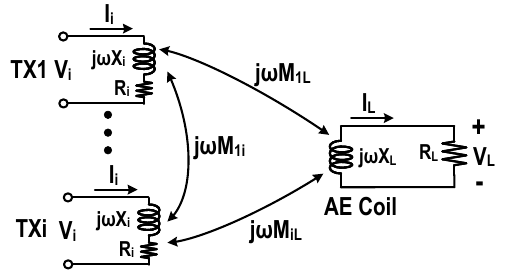}
\vskip -1ex
\caption{
{Simplified equivalent model for multiple TX coil array and AE coil with an ideal AC load $R_L$.}}
\label{multi_coil_concept}
\end{figure}

Alternatively, we realize that the full 6-DoF information is not necessary for guiding the field steering and propose the Active Echo (AE) coupling sensing technique for multi-coil adaptive control. The idea is to actively transmit an echo signal from the implant and use the received signal's amplitude and polarity on the external TX to directly determine the current allocation. 
{Using the converse magnetoelectric effect of ME~\cite{Mehdi_converse_ME} as the Active Echo transmitter was an initial exploration. However, its complex and non-reciprocal energy conversion process leads to a much lower misalignment sensitivity and signal strength, as shown in Fig.~\ref{AE_ME_revise}. These properties necessitated a higher-performance receiver chain and a calibration method to accurately locate the implant and control multiple coils, significantly complicating system analysis and loop closure. This calibration is also required for each ME film.} 
Here, our workaround is to add a tiny auxiliary coil on the implant to transmit the AE signal. By leveraging the reciprocity of coil coupling~\cite{haus_electromagnetic_1989}, we can directly obtain the optimal current allocation among coils, without calculating or estimating the pose of the implant. 
Next, we will explain this principle by modeling the AE coil and the TX array as in Fig.~\ref{multi_coil_concept}. Based on the Kirchhoff's law, one can obtain \cite{qiu_678-mhz_2022, zhang_efficiency_2016},

\begin{equation}
\resizebox{0.9\columnwidth}{!}{
$ 
\left[\begin{array}{cccc}
R_1+j X_1 & j \omega M_{12} & \cdots & j \omega M_{1 L} \\
j \omega M_{21} & R_2+j X_2 & \cdots & j \omega M_{2 L} \\
\vdots & \vdots & \ddots & \vdots \\
j \omega M_{n 1} & j \omega M_{n 2} & \cdots & j \omega M_{n L} \\
j \omega M_{L 1} & j \omega M_{L 2} & \cdots & R_L+j X_L
\end{array}\right] \cdot\left[\begin{array}{c}
I_1 \\
I_2 \\
\vdots \\
I_n \\
I_L
\end{array}\right]=\left[\begin{array}{c}
V_1 \\
V_2 \\
\vdots \\
V_n \\
0
\end{array}\right]
$ 
}
\label{matrix_power}
\end{equation}
where $R_i$ and $X_i$ represent the resistance and reactance of each resonance tank. $M_{1i}$ is the mutual inductance between $TX_1$ and $TX_i$, and $M_{iL}$ is the mutual inductance between TX$_i$ and RX. 
We can easily derive the current on the receiver side as,
\begin{equation}
    I_L=-\frac{j \omega \left(M_{L1} I_1+M_{L2} I_2+\cdots+M_{Ln} I_n\right)}{R_L+j X_L}
    \label{received power}
\end{equation}

{The wireless power transfer efficiency (PTE) in this paper is defined by the ratio of total AC power consumption of the TX coils and the maximum AC power received by the RX with an optimal loading,}
\begin{equation}
\begin{aligned}
\eta_{w t} & =\frac{P_{\text {Recv}}}{P_{\text {Loss}}+P_{\text {Recv}}} =\frac{1}{1+{\frac{P_{\text {Loss}}}{P_{\text {Recv}}}}}
\end{aligned}
\label{eq_3}
\end{equation}

Now, the optimization goal is to minimize power loss and maximize efficiency, which can be achieved by maximizing ${\frac{P_{\text {Recv }}}{P_{\text {loss }}}}$ according to 
(\ref{eq_3}). 

\begin{equation}
\frac{P_{\text{Recv}}}{P_{\text{loss}}}=\frac{R_L\left|I_L\right|^2}{R_1\left|I_1\right|^2+R_2\left|I_2\right|^2+\cdots+R_n\left|I_n\right|^2} 
\label{eq_max}
\end{equation}

Using (\ref{received power}) to represent $I_L$, one can get,

\begin{equation}
\resizebox{0.9\columnwidth}{!}{
$ \frac{P_{\text{Recv}}}{P_{\text{loss}}}=\frac{R_L \omega^2}{R_L^2+X_L^2} \cdot \frac{\left|M_{L1} I_1+M_{L2} I_2+\cdots+M_{Ln} I_n\right|^2}{R_1\left|I_1\right|^2+R_2\left|I_2\right|^2+\cdots+R_n\left|I_n\right|^2} $
}
\label{eq_4}
\end{equation}

Here, only the second term depends on the receiver state, one can apply Cauchy-Schwarz Inequality to its numerator,
\begin{equation}
\resizebox{1\columnwidth}{!}{
$\begin{aligned}
& \left|M_{L1} I_1+M_{L2} I_2+\cdots+M_{Ln} I_n\right|^2 \\
& =\left|\frac{M_{L1}}{\sqrt{R_1}} \sqrt{R_1} I_1+\frac{M_{L2}}{\sqrt{R_2}} \sqrt{R_2} I_2+\cdots+\frac{M_{Ln}}{\sqrt{R_n}} \sqrt{R_n} I_n\right|^2 \\
& \leq\left(\frac{M_{L1}^2}{R_1}+\frac{M_{L2}^2}{R_2}+\cdots+\frac{M_{Ln}^2}{R_n}\right)\left(R_1\left|I_1\right|^2+R_2\left|I_2\right|^2+\cdots+R_n\left|I_n\right|^2\right)
\end{aligned}$
\label{inequality}
}
\end{equation}

By substituting (\ref{inequality}) into (\ref{eq_4}), we get,

\begin{equation}
\resizebox{0.9\columnwidth}{!}{
$\begin{aligned}
\frac{P_{\text {Recv }}}{P_{\text {Loss }}} \leq \frac{R_L \omega^2}{R_L^2+X_L^2}\left(\frac{M_{L1}^2}{R_1}+\frac{M_{L2}^2}{R_2}+\cdots+\frac{M_{Ln}^2}{R_n}\right)
\end{aligned}$
}
\end{equation}

where the equality exists only when $\frac{M_i}{\sqrt{R_i}} = \sqrt{R_i} I_i$. If we assume all transmitters are identical and thus $R_i$ of all coils are the same, the equality condition becomes, 

\begin{equation}
\frac{I_1}{M_{L1}}=\frac{I_2}{M_{L2}}=\cdots=\frac{I_n}{M_{Ln}}
\label{I_ratio}
\end{equation}

This is the optimal condition to achieve maximum WPT efficiency. Since ${M_{iL}}$ equals to ${M_{Li}}$ in an electromagnetic induction system \cite{haus_electromagnetic_1989}, the mutual inductance is proportional to the coupling coefficient. So (\ref{I_ratio}) can be further simplified to,

\begin{equation}
    \frac{I_1}{k_1}=\frac{I_2}{k_2}=\cdots=\frac{I_n}{k_n}, \quad k_n \propto M_n
    \label{K_ratio_eq_I_ratio}
\end{equation}

\begin{figure*}[!t]
\centering
\includegraphics[width=\linewidth]{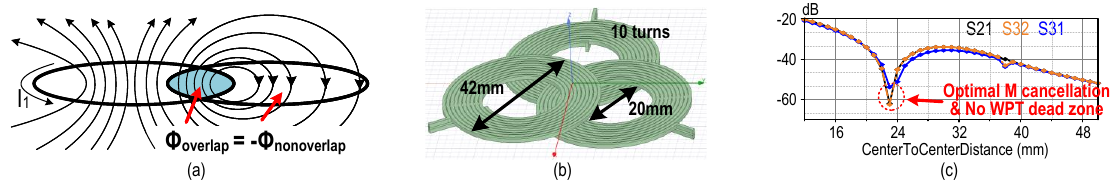}
\vskip -1ex
\caption{(a) Two coils with mutual inductance cancellation; (b) Three coils modeled in HFSS; (c) S parameter simulation results of the three-coil array in HFSS.}
\label{M_cancel}
\end{figure*}


Eq.~(\ref{K_ratio_eq_I_ratio}) reveals our control strategy for AE-enabled multi-coil TX. 
The system achieves peak efficiency when the ratio of currents allocated to TX coils matches the ratio of their coupling coefficients with the receiver coil. This condition also implies that the magnetic field perpendicular to the AE coil reaches its maximum. In addition to the amplitudes of coil currents, we also need to control their phases. 
{Since the wavelength of the ME film's resonance frequency (340kHz) is approximately 882m, which is significantly larger than the distance between TX and RX in our application, there is negligible phase difference between the AE signal received by each channel.} Consequently, we can only observe identical or opposite polarities in AE sensing. On the other hand, when multiple TX coils are involved in magnetic field generation, the field induced by each coil will see a negligible phase difference within the space of interest. Therefore, we only need to set the current of each coil to two opposite phases to most efficiently add up or cancel their contributions for steering purposes. Applying any other phase shifts will only consume higher TX power for the same result. For these reasons, we are only concerned about sensing and controlling two opposite phases and we can sense the polarities in AE sensing and use them directly as the optimal phase for each TX coil.  
Finally, by aligning the long axis of an ME film (i.e., its ideal field direction) with the normal direction of the AE coil, we ensure the ME transducer gets the strongest possible magnetic field along its long axis, under a given total TX power budget.

\subsection{Mutual Inductance Cancellation in Multi-Coil TX}

AE effectively addresses the weak coupling challenge, but the mutual inductance between TX coils is another issue. To better analyze this, we calculate the driver voltage and current on the TX side in a multi-coil system form Eq.~(\ref{matrix_power}).

\begin{equation}
    V_{i}=R_{i} I_{i}+\sum_{k \neq i} j \omega M_{ i k} I_{k}
\end{equation}
where $V_{i}$, $R_{i}$ are the output voltage of i-th driver and the impedance of the resonance tank, respectively. $M_{i k}$ represents the mutual inductance between i-th TX and k-th TX. In our system, since the $I$ varies for each transmitter, the contribution of the second term differs for each TX, leading to a current-dependent resonant frequency shift and reducing the effective AC current in the TX tank. In an inductive coupling system, one can adjust the operating frequency to compensate for the degradation of resonant current. However, for the ME film, the carrier frequency has to match the ME's acoustic resonant frequency (340kHz) to achieve high efficiency.
Therefore, we take a principled approach to optimize the relative position of multiple coils to minimize their mutual inductance. 

{We first explain the principle with two overlapping coils shown in Fig.~\ref{M_cancel}~a, where the mutual inductance is,
\begin{equation}
\begin{aligned}
    M_{21}&=M_{12}  =\frac{\Phi_{2,total}}{i_1} = \frac{\Phi_{overlap}+\Phi_{nonoverlap}}{i_1} 
\end{aligned}
\end{equation}
where $M$ is the mutual inductance between two coils. Apparently, matching $\Phi_{overlap}$ and $\Phi_{nonoverlap}$ values with opposite signs will cancel the total flux and minimize their mutual inductance. Since $\Phi = \int_{S}{} \mathbf{B} da$, we can adjust the distance between the two coils to change the overlapping area and achieve mutual inductance cancellation. The specific distance can be determined experimentally using HFSS simulation and physical measurements. By applying the same criterion to three coils, we can extend the field control ability to 6 DoF and ensure that the mutual inductance is canceled for all pairs of coils, resulting in the 3-coil array in our system (Fig.~\ref{M_cancel}~b). }

Simulation results in Fig.~\ref{M_cancel}~c indicate that the mutual inductance in this 3-coil design is even smaller than parallel coils with no overlapping. 
Moreover, the overlapping coil placement avoids regions with overly small magnetic field strength despite coil current optimization, such as the gap regions among non-overlapping coils.

\section{Circuit and System Implementation}

\begin{figure*}[!t]
\centering
\includegraphics[width=\linewidth]{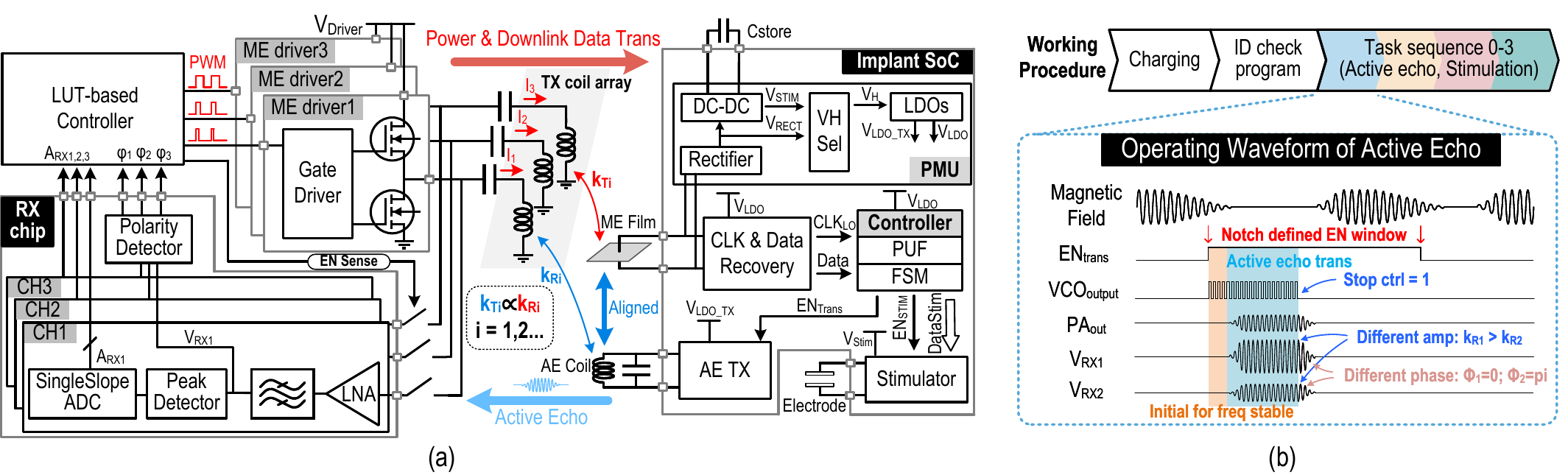}
\vskip -1ex
\caption{(a) Block diagram of the AE-enabled omnidirectional ME WPT platform; (b) Operation flow of our implant and the waveforms of AE phase.}
\label{system_overview}
\end{figure*}

\subsection{System Overview}
Fig.~\ref{system_overview} illustrates the complete omnidirectional WPT system for millimetric ME implants in this work. 
First, we use three off-the-shelf GaN drivers to drive three transmitter coils. The coils meet the proposed floor plan to cancel the mutual inductance in between. This phase initially powers on the implant and charges its storage capacitor.
Then, the implant recovers its CLK from the carrier frequency and decodes the downlink instructions in the central controller. To overcome the trade-off between the downlink data rate and transducer quality factor, we utilize a time-domain notch-based modulation scheme\cite{yu_wireless_2022}. 
{In the meantime, the ME-induced voltage is rectified to $V_{rect}$ and then converted by the DC-DC. The system selects the higher voltage between $V_{rect}$ and $V_{stim}$ to power the LDOs and the rest of the system, ensuring the working stability, even if there are fluctuations in $V_{rect}$ due to misalignment.}
{After establishing the downlink communication, the central controller first checks the device ID before programming the implant based on the downlink and executing tasks sequentially. 
We use an on-chip Physical Unclonable Function (PUF) \cite{Dai_PUF} to produce the device ID with minimum area and power, eliminating the need for non-volatile memories and programming. } 
Our prototype includes two available tasks, programmable stimulation and AE sensing.


When the AE task is enabled, the TX driver turns off all the power transistors and enters a high-impedance mode, while the AE TX in the implant is activated to send out a few cycles of a single tone through the AE coil. 
{Thanks to the low turn-off leakage of the driver, there is no DC drift on the receiver input when the driver turns off. Therefore, the AE signal is captured by the external TX coil with high fidelity and fed into RX chip to determine the amplitude and polarity in each channel.}
Finally, the central controller in TX uses this information to generate PWM signals for the coil drivers.
{We store the transfer curve between the duty cycle and output current as a look-up table (LUT) in the MCU because it's not perfectly linear. When the MCU receives the amplitude and polarity information from the RX, it first adjusts the output current to be proportional to the AE amplitude, and the current polarity aligns with AE output. To ensure safety, the total power budget is fixed. This means $I_1^2+I_2^2+I_3^2$ is constant, allowing us to determine the required current value for each channel. Using the look-up table, the MCU then generates the corresponding PWM waveforms.}

This procedure is performed periodically to track the implant's movements. 
{The frequency to repeat AE sensing is called the AE activation frequency, representing how fast the system updates the TX current configurations. This frequency is programmable and application-dependent. For example, in one of our targeted scenarios—cardiac pacemaker—using a 20Hz sampling rate, around 10 times the heart rate, is sufficient to track heartbeat-caused implant movement. 
During each AE sensing phase, the ring voltage-controlled oscillator (VCO) in the AE TX is activated 8 cycles in advance to stabilize. Considering the 16 cycles of AE transmission and ME transducer's ring up/down time, each AE operation takes less than 100us. The interruption to ME power transfer by AE sensing is thus negligible.}

To demonstrate the proposed WPT platform and further enhance system integration and portability, we developed two chips: an implantable bio-stimulator SoC with a technical emphasis on a three-level low-power AE TX,
and a three-channel AE RX, including a compact, low-noise analog front-end, single slope ADC, and fast polarity detector. 
\subsection{AE Transmitter in the Implant SoC}

\begin{figure}[t!]
\centering
\includegraphics[scale = 0.66]{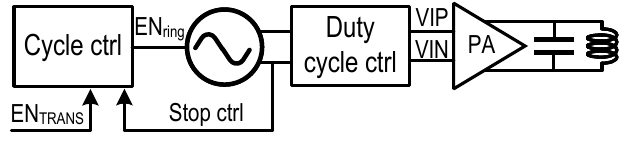}
\vskip -2ex
\caption{Top level structure of proposed AE transmitter.}
\vskip -2.5ex
\label{PA_TOP}
\end{figure}

Fig.~\ref{PA_TOP} shows the low-power AE TX circuits to drive the AE coil. 
The cycle controller is synchronized with the external device through the task trigger signal $ENtrans$, which is handled by the implant's central controller. When $ENtrans$ goes high, the cycle controller activates the ring VCO, counts the number of output cycles, and stops the VCO after a programmable number of cycles. H-bridge PA is chosen for its high power efficiency with no static power loss. 
Using a square wave to drive the H-bridge PA is a straightforward choice.
{However, the square wave includes undesirable harmonics, while only the fundamental tone injected into the LC tank, behaving as a bandpass filter, is useful for AE coupling sensing. To save the power, }
we adopt the multi-level PA concept to better mimic a sinusoidal signal \cite{fritzin_design_2012-1}, aiming to reduce the driver's turn-on time and its power without compromising the amplitude of the fundamental tone.

\begin{figure}[t!]
\centering
\subfloat[]{\includegraphics[width = 0.9\linewidth]{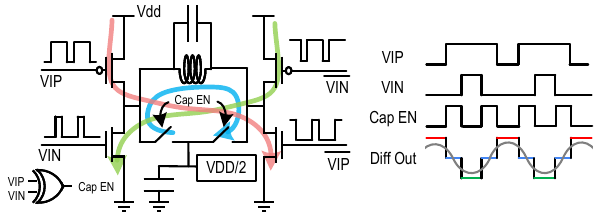}%
\label{FIg_8_PA_a} 
}\\
\subfloat[]{\includegraphics[width = 0.9\linewidth]{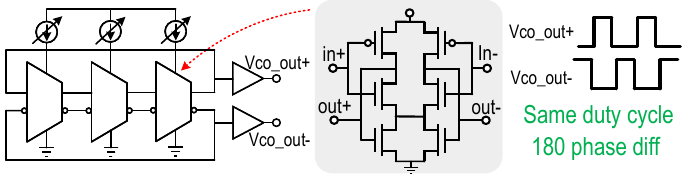}%
\label{FIg_8_PA_b}
} \\
\subfloat[]{\includegraphics[width = 0.9\linewidth]{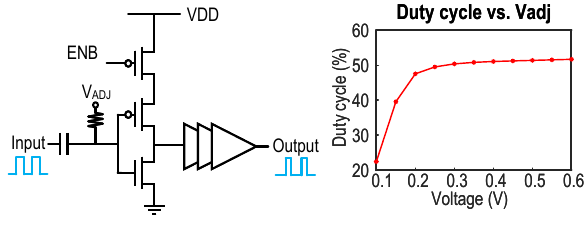}%
\label{FIg_8_PA_c}
}
\caption{(a) 3-level low-power PA and its operation waveforms; (b) VCO design; (c) Analog duty cycle control circuits and simulated performance.}
\label{FIg_8_PA}
\end{figure}

For simplicity and compactness, we implemented a three-level PA, by controlling the duty cycle and adding a non-overlapping phase to the VCO’s differential outputs. 
As shown in Fig.~\ref{FIg_8_PA_a}, the H bridge in the low-power PA is completely turned off when the output is at the blue-colored middle voltage. During this phase, the outputs are shorted to a pre-charged reference voltage, forming a current loop. Theoretically, there is no power consumption in this phase.

To produce the symmetrical VCO outputs with the same duty cycle but at 180\textdegree{} phase shift, we designed a differential ring oscillator with a self-biased current tail~\cite{musa_compact_2014-1,wang_36nw_2023}, as shown in Fig.~\ref{FIg_8_PA_b}. 
The tail design acts as a current source without requiring additional gate biasing and behaves the same no matter with branch is on, ensuring identical common-mode output voltage and duty cycle for the differential outputs. 
The VCO outputs are then sent to the duty cycle control (DCC) circuit, which adjusts the duty cycle by changing the input biasing of the AC-coupled inverter, as shown in Fig.~\ref{FIg_8_PA_c}. A 3-bit voltage digital-to-analog converter (DAC) allows programming this biasing voltage from 0.1 to 0.5V.

\subsection{Multi-Channel AE Receiver (RX)}
On the external hub side, conventionally, a separate pickup coil is used to capture the uplink signal\cite{wang_design_2005,yu_wireless_2022}. However, these two coils are coupled and difficult to align, which will degrade the accuracy of coupling coefficient sensing in the AE scheme. Meanwhile, this two-coil topology doubles the number of coils and complicates the implementation. 

\begin{figure}[t!]
\centering
\includegraphics[scale = 0.58]{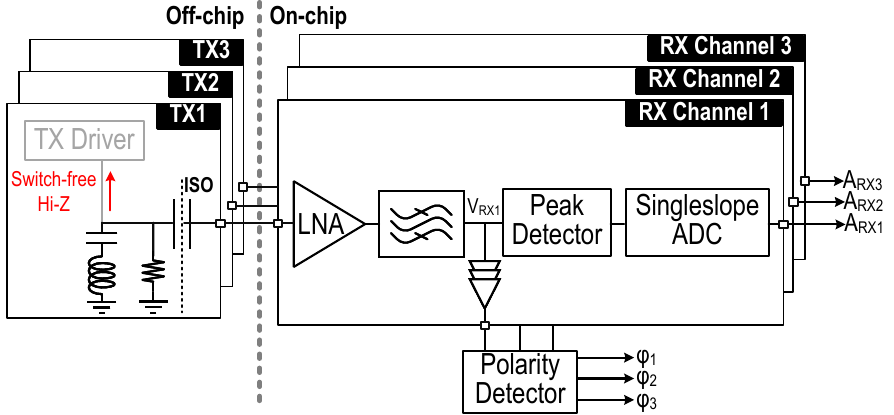}
\vskip -1ex
\caption{Architecture of the AE receiver.}
\label{RX_TOP}
\end{figure}

Fig.~\ref{RX_TOP} shows the architecture of the AE RX. We reuse the same coils for powering and sensing at different frequencies and times. The magnetic field for power transfer is at 340kHz, while the AE signal is designed to be roughly 1MHz. This scheme makes it vastly easier to scale up the number of coils for wider coverage. 
{The AE sensing phase is triggered by the external hub, which ensures the synchronization. When the implant transitions to the AE task, the TX driver enters the high-impedance mode so that the AE signal captured by the TX coils can flow into the receiver chain without attenuation.}
After low noise amplification and band-pass filtering, the signal $V_{RX}$ is sent to the peak detector and digitized by a single slope ADC, in the meantime, this signal is further amplified to rail-to-rail and sent to a digital polarity detector.

\subsubsection{Low Noise Analog Front End}
\begin{figure}[t!]
\centering
\includegraphics[scale = 0.65]{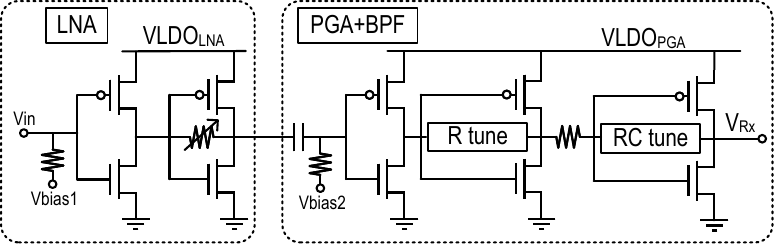}
\vskip -1ex
\caption{Analog front-end circuits in the AE receiver.}
\label{RX_AFE}
\end{figure}

An inverter-based LNA and PGA are designed for good noise and programmable gain control, As shown in Fig.~\ref{RX_AFE}. 
{We overdesigned the noise performance of the receiver chain to leave the SNR margin in the worst case, ensuring the AE served range is larger than the ME power range.}
The low gain mode ensures that the output of the LNA and PGA will not saturate when the implant is very close to the TX and the TX coils capture a strong AE signal. In this LNA topology, inserting configuration switches at the first $Gm$ stage to control the total gain will either occupy a large area to reach a low insertion resistor or degrade the total $Gm$. Therefore, we only change the transimpedance in the second stage of LNA, keeping the total $Gm$ constant. This leads to a near-constant input referred noise under different gain configurations. The first stage of PGA uses the same structure with a relaxed noise requirement, and the second stage replaces the OTA with a resistance to handle a large voltage swing and maintain linearity. Bandpass characteristic consists of two passive high pass filters and an active RC low pass filter to suppress the interference from TX and narrow the noise band.
LNA and PGA are biased by duplicated inverter structures to track PVT variations.
One drawback of the inverter topology is poor power supply rejection (PSR), which compromises noise performance and isolation between adjacent channels. To mitigate this issue, we employ local LDOs for LNA and PGA individually in each channel. This design also ensures stability of the AFE, by breaking the positive feedback loop from the LNA's supply to the front-end's output.


\begin{figure}[t]
\centering
\subfloat[]{\includegraphics[scale = 0.7]{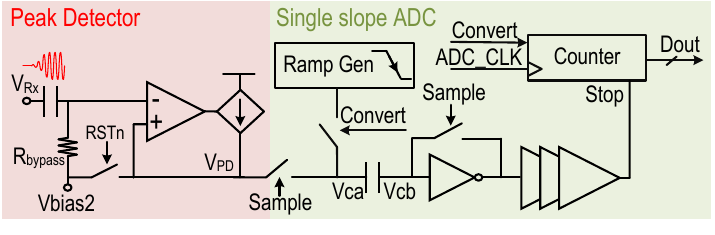}%
\label{RX_ADC_a}
}\\
\subfloat[]{\includegraphics[scale = 0.65]{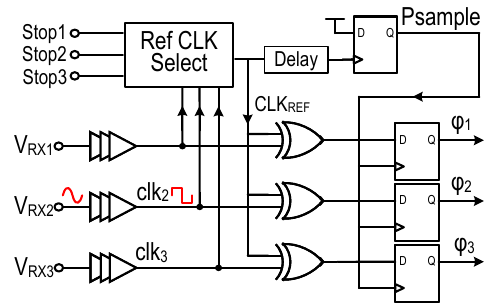}%
\label{RX_ADC_b}
}
\hfill
\subfloat[]{\includegraphics[scale = 0.67]{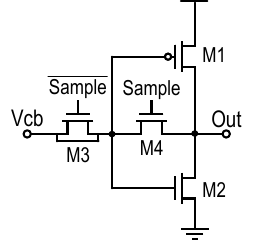}%
\label{RX_ADC_c}
}
\caption{(a) Peak detector and single slope ADC structure; (b) digital polarity detector; (c) inverter-based comparator in single slope ADC.}
\label{RX_ADC}
\end{figure}

\subsubsection{Peak Detector and Data Converter}
The amplitude digitization circuitry is shown in Fig.~\ref{RX_ADC_a}. The peak detector tracks the peak voltage of $V_{Rx}$ and stores it on the capacitor. Meanwhile, the single-slope ADC (SS-ADC) performs auto-zero on its comparator. 
The sampler waits a few cycles for the AE signal to ramp up and stabilize,  before starting the SS-ADC.
During data conversion, a reverse ramp signal from VDD to GND is supplied to the left plate of the sampling cap ($V_{ca}$), and the counter starts. When $V_{cb}$ reaches the threshold voltage of the inverter, the counter stops and outputs its value. Being on the external side, this counter is clocked by a high-quality off-chip clock source. The inverter-based comparator is shown in Fig.~\ref{RX_ADC_c}, where M3 is a dummy switch to compensate for the charge injection induced by M4.

\begin{figure}[t!]
\centering
\includegraphics[scale = 0.72]{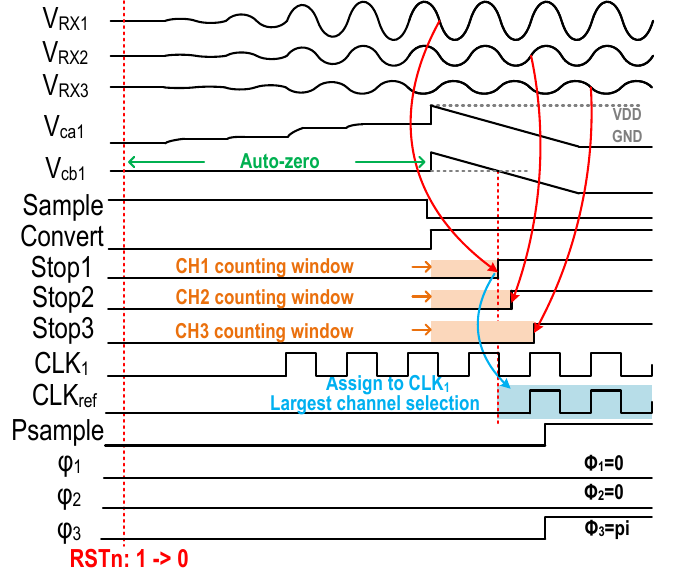}
\caption{Operation waveforms of the AE receiver chip.}
\vskip -2ex
\label{RX_waveform}
\end{figure}

\subsubsection{Polarity Detector}
As discussed in \S II, we need to sense the polarities of $V_{Rx}$ in all channels, in addition to their amplitudes. Determining the polarities of these iso-frequency signals can be efficiently realized by performing XOR logic on two amplified rail-to-rail $V_{Rx}$ signals. Inputs with identical polarities produce 0, while inputs with opposite polarities produce 1. 
%
%
Instead of comparing all pairs, we can select any channel as a reference clk and determine the relative polarities of all other channels. This is feasible because we only need the relative polarities for multi-coil control. 
However, sometimes not all channels receive a strong enough AE signal to be reliably amplified into a digital signal. 
Thus, selecting the channel receiving the strong AE signal as the reference clock is the most reliable way for the polarity detection (Fig.~\ref{RX_ADC_b}). To quickly find the strongest channel, we simply utilize a reverse ramp for SS-ADC so that higher amplitude signals take less time to finish the conversion. This design contributes to reliable and fast polarity detection, minimizing the length of the AE sequence and thus saving the power of the implant. 

Fig.~\ref{RX_waveform} shows sample waveforms to illustrate the operations of the AE RX. Here, channel 1 receives the strongest signal. After the reset signal is released, the peak detector first samples the peak value of each channel, and the right plate of the sampling cap is auto-zeroing simultaneously. Then, the ADC starts conversion and concludes first for Channel 1. As a result, the reference selection circuit immediately uses channel 1 as the reference clock for polarity detection. Then the XOR output will be sampled as each channel's polarity relative to channel 1. After the amplitude conversions of all channels finish, all six measurements are sent to the controller for optimizing the TX coil configuration.
{It is worth noting that we only need the relative polarities among channels, rather than the absolute values. Thus, even if two or three channels stop at almost the same time and the clock selection circuit makes a mistake due to noise and variations, the sensed relative polarity among channels is sufficient to ensure AE's functionality. }

\section{Experimental Results}
\begin{figure}[t]
\centering
{\includegraphics[width=.85\linewidth]{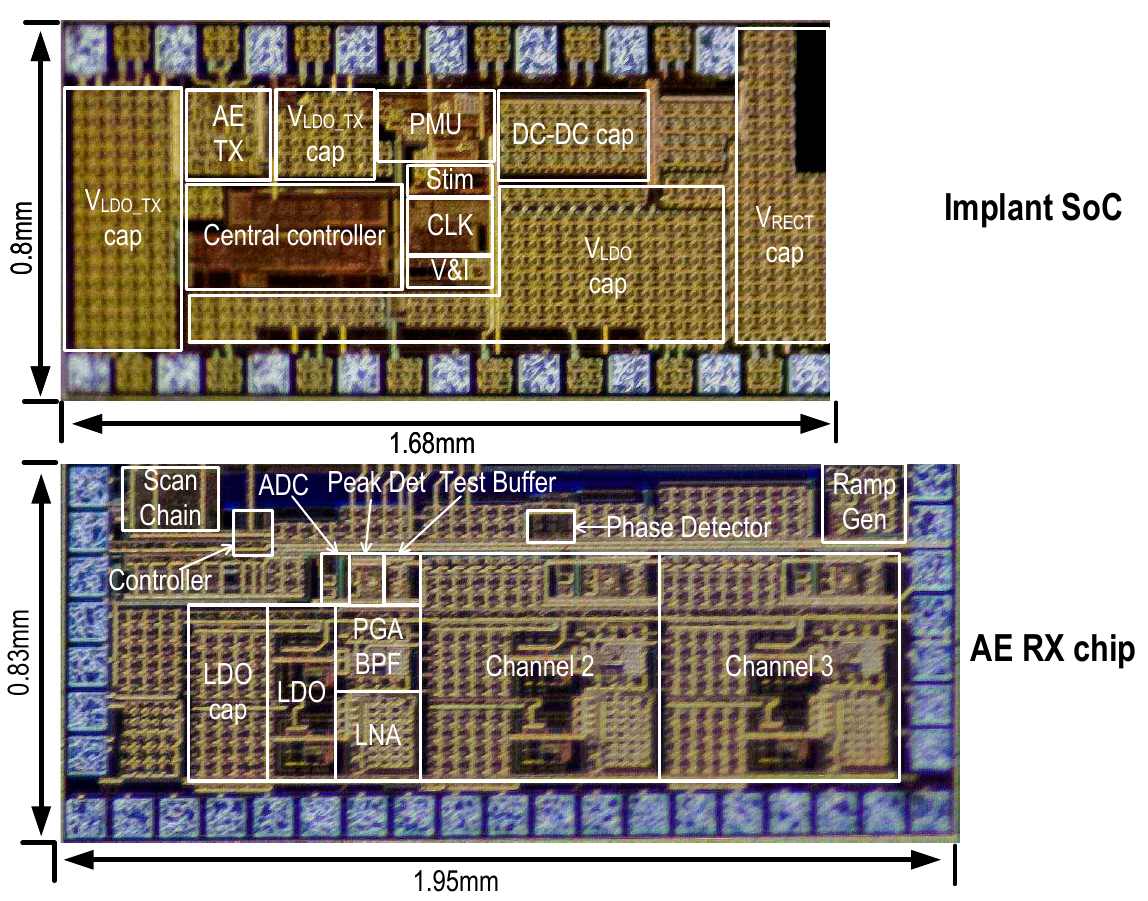}%
} \\
\hfil
\vskip -2ex
\caption{Chip micrographs of the implant's SoC and the three-channel AE RX chip. }
\label{die_photo}
\end{figure}

The implant's SoC and three-channel AE RX chip are fabricated in TSMC 180nm CMOS
technology with respective areas of 1.34$mm^2$ and 1.62$mm^2$ (see Fig.~\ref{die_photo}). We first test the performance of the individual chips and then integrate them into a complete wireless system for evaluation.

\subsection{Active Echo (AE) Transmitter Testing}
\begin{figure}[t!]
\centering
\includegraphics[width=.95\linewidth]{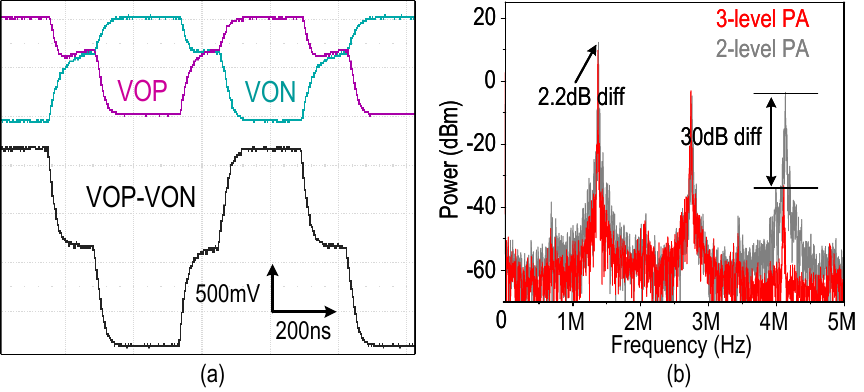}
\vskip -2ex
\caption{3-level low power PA testing results without LC loading: (a) transient waveform; (b) differential output spectrum.}
\label{Test_PA}
\end{figure}

\begin{figure}[t!]
\centering
\includegraphics[scale = 0.6]{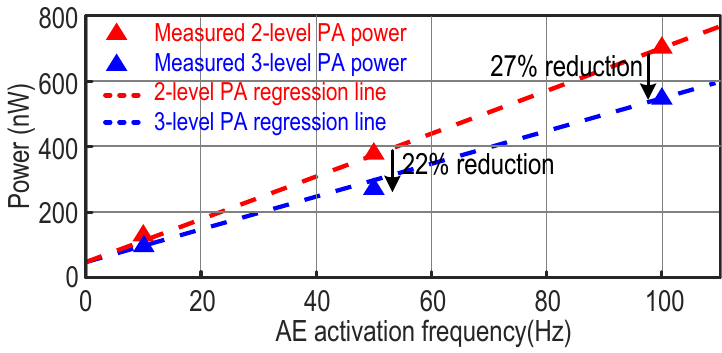}
\caption{Measured power consumption of AE TX with 16 running cycles at different activation frequencies.}
\vskip -2ex
\label{Test_PA_C}
\end{figure}

When a low-impedance LC loading is presented to the differential output of the AE TX, the voltage swing decreases, and the LC loading reshapes the output waveform. To better assess the output waveform and spectrum of the three-level PA, we first remove the LC loading at the TX output to obtain the measurement results in Fig.~\ref{Test_PA}~(a,b).
The TX is powered by a 1V LDO separately to ensure better isolation from other circuits during operation. Here, the clock duty cycle produced by DCC is set to around 30\%. As designed, the output voltage briefly stays near half VDD when all transistors of the PA are off. As shown by the black line in Fig.~\ref{Test_PA}~a, the differential output of the AE TX is a symmetrical three-level waveform that better mimics a sinusoidal waveform than a basic square wave. 
By adjusting the biasing voltage of the DCC block, we can tune the duty cycle to achieve better harmonic suppression. In our testing, the output spectrum achieves an approximately 30dB 3rd-order harmonic reduction with only a 2.2dB fundamental tone loss at around 30\% duty cycle. 
Since the AE TX operates for only a few cycles after activation, its power consumption is greatly amortized, and the average power remains minimal. As shown in Fig.~\ref{Test_PA_C}, the power consumption linearly increases with AE activation frequency, where the multi-level technique reduces power by 27\% at 100Hz. By extending the regression lines to 0Hz, we can estimate that both 2- and 3-level PAs have a similar static power of roughly 50nW. 


\subsection{Three-Channel AE Receiver Testing}
\begin{figure}[t!]
\centering
\includegraphics[width=.95\linewidth]{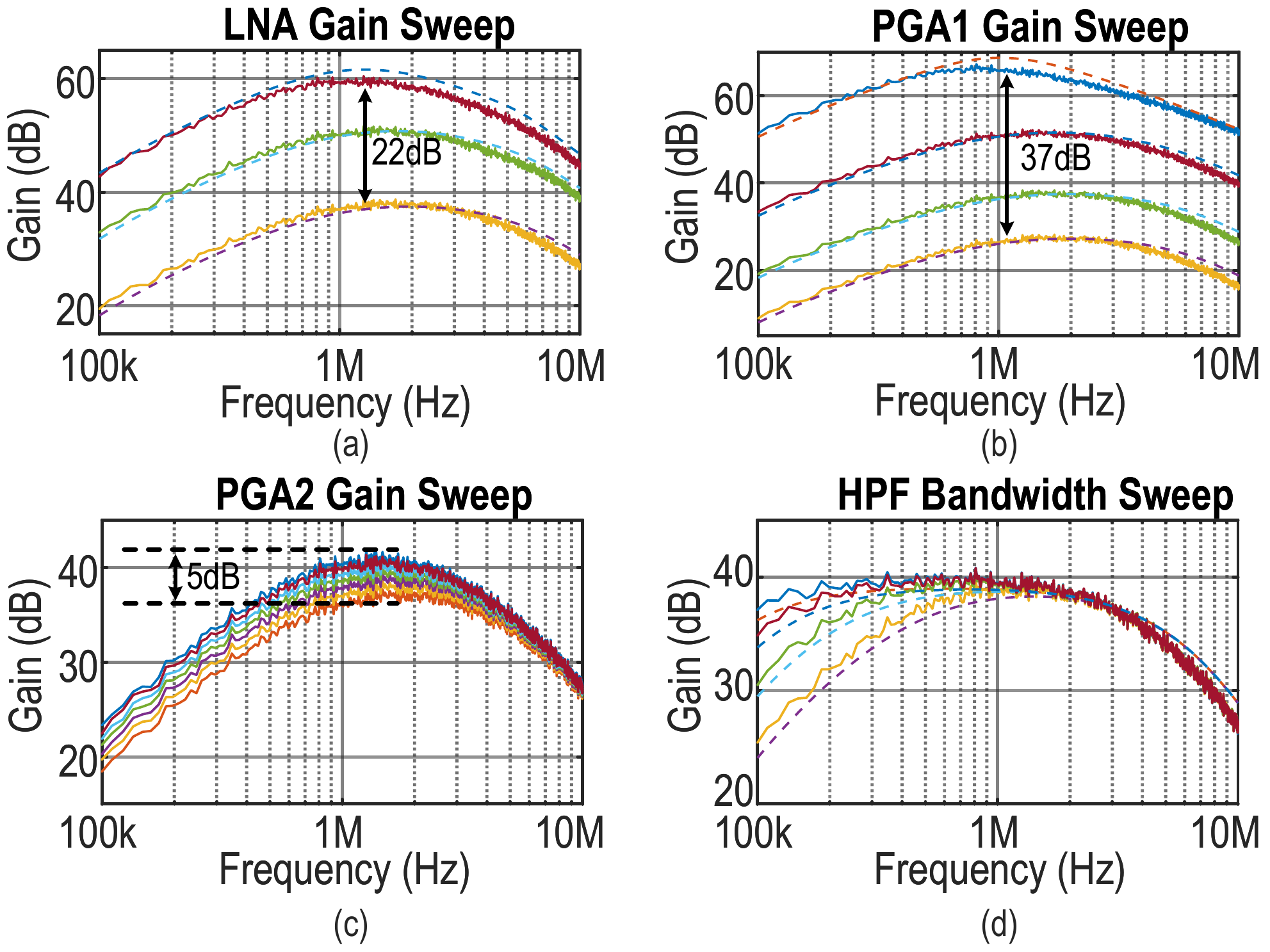}
\vskip -2ex
\caption{
{Measured gain and bandwidth across different configurations of the AE RX front-end (solid lines: measurement; dashed lines: simulation).}}
\label{AFE_gain}
\end{figure}

We first measure the gain and bandwidth of a single AFE channel with an impedance analyzer, as shown in Fig.~\ref{AFE_gain}. 
LNA achieves a 22dB tuning range by configuring the resistor within the TIA stage. The first-stage PGA has the same gain control strategy with a 37dB tuning range. To compensate for PVT variations and balance the three channels, PGA2 offers a 5dB tuning range with less than 1dB step, as shown in Fig.~\ref{AFE_gain}~c. Fig.~\ref{AFE_gain}~d shows four programmable high-pass cut-off frequencies from 100kHz to 600kHz, filtering low-frequency interference and noise. 
Across various configurations of the AFE, measurements (solid lines in Fig.~\ref{AFE_gain}) closely match the simulations (dashed lines in Fig.~\ref{AFE_gain}). The slight mismatch at high gains is due to the limited SNR of the signal source. 

\begin{figure}[t!]
\centering
\includegraphics[width=.92\linewidth]{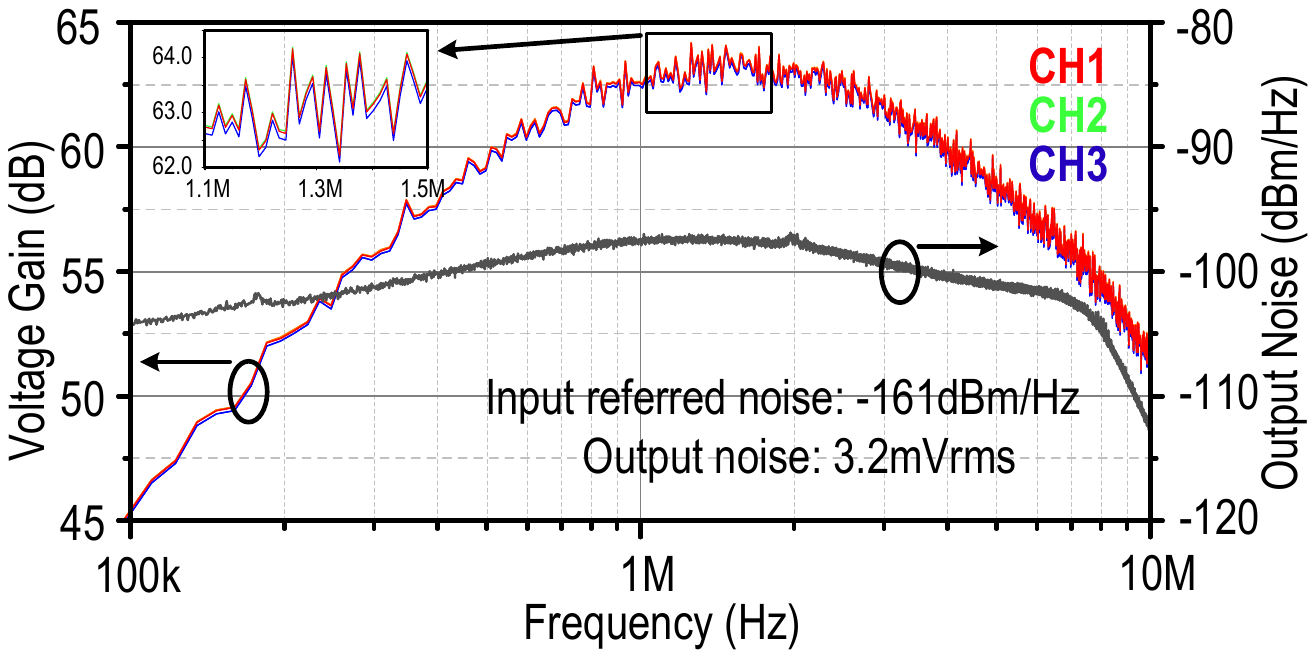}
\vskip -2ex
\caption{
{Measured AFE gain of 3 channels and output noise spectrum of channel 1.}}
\label{gain&noise}
\end{figure}

For system demonstrations, we set the channel gain to $\sim$63dB at 1.5MHz to achieve the largest $V_{Rx}$ swing (around 1.5V) with perfectly aligned TX and RX. 
{Fig.~\ref{gain&noise} shows the gain responses of all three channels and the noise performance of channel 1 under this configuration. The zoom-in view indicates the silicon mismatch after calibration is less than 0.2dB. The fluctuation is caused by limited equipment calibration accuracy.}
The input-referred noise is around -161dBm/Hz, calculated from the output noise floor and the channel gain. 

\begin{figure}[t!]
\centering
\includegraphics[width=.9\linewidth]{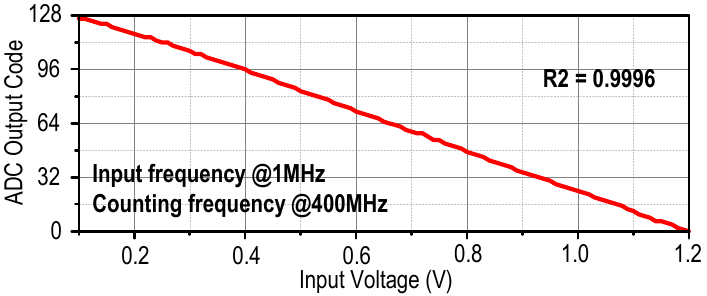}
\vskip -2ex
\caption{Measured conversion linearity of peak detector and ADC.}
\label{ADC}
\end{figure}

Next, we test the peak detector and ADC together as they are integrated. A 1MHz sinusoidal signal is input into the peak detector, which converts it to a DC value. The DC value is then digitized by a cascaded single-slope ADC using a 400 MHz external clock. 
Fig.~\ref{ADC} shows the linearity of the output digital code when the input voltage sweeps from 0.1 to 1.2V. 
The $R^2$ value of the transfer curve is 0.996. The linearity is primarily restricted by the gain error variations of OpAmps in the ramp signal generator and the peak detector. The linearity of the peak detector is the more dominant factor since the speed of the ramp signal is much lower than the desired RX signal.

The power measured from the single channel AE RX is 18.65mW at a 1.8V supply voltage, comprising 10.2mW LNA power, 3.5mW PGA power, 2.5mW ADC power, and 2.45mW from clock path and digital circuity. The LNA consumes more than half of the total power to reduce input-referred noise for weak AE signal sensing, which is overdesigned to guarantee a sufficient SNR when the implant is positioned a little further away from the TX coils.

\subsection{Mutual Inductance Cancelled TX Coil Array}
\begin{figure}[t!]
\centering
\includegraphics[width=.95\linewidth]{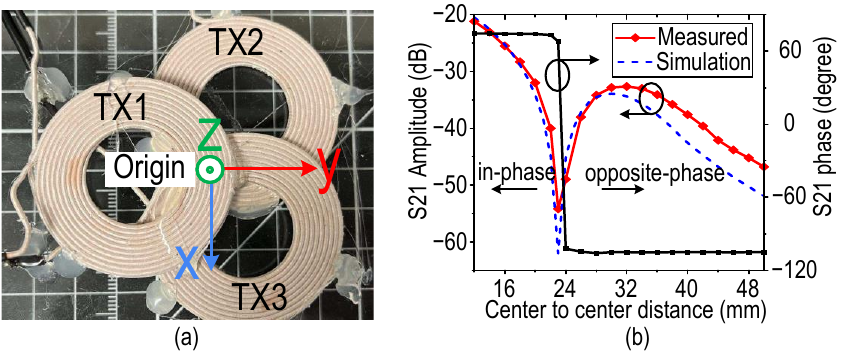}
\vskip -2ex
\caption{(a) A picture of the 3-coil TX array, marked with our coordination system; (b) S21 results for two coils in measurement and HFSS simulation.}
\label{M_canel_results}
\end{figure}

We assemble off-the-shelf coils made with Litz wire into a 3-coil TX array, as shown in Fig.~\ref{M_canel_results}~a. Each coil has its own compensation capacitor to match the resonant frequency of our ME film (around 340kHz) for optimal PTE. 
%
The optimized placement of the three coils minimizes the mutual inductance. When the center-to-center distance of the coils is 24mm, TX coils present minimum coupling. The abrupt phase change at this point also indicates that the amplitude of S21 crosses zero (Fig.~\ref{M_canel_results}~b). The measurement also closely matches the simulation results. 

Fig.~\ref{M_canel_results}~a also illustrates the coordination system used in our measurements, with the origin set to the geometric center of the three coils. We use brackets, (x,y,z), to notate the location of the implant in the unit of millimeters and curly brackets, {x,y,z}, to denote the direction of the implant, i.e., the direction of the long axis of the ME film. 

\subsection{System In-Vitro Testing}
\begin{figure}[!t]
\centering
{\includegraphics[scale = 0.36]{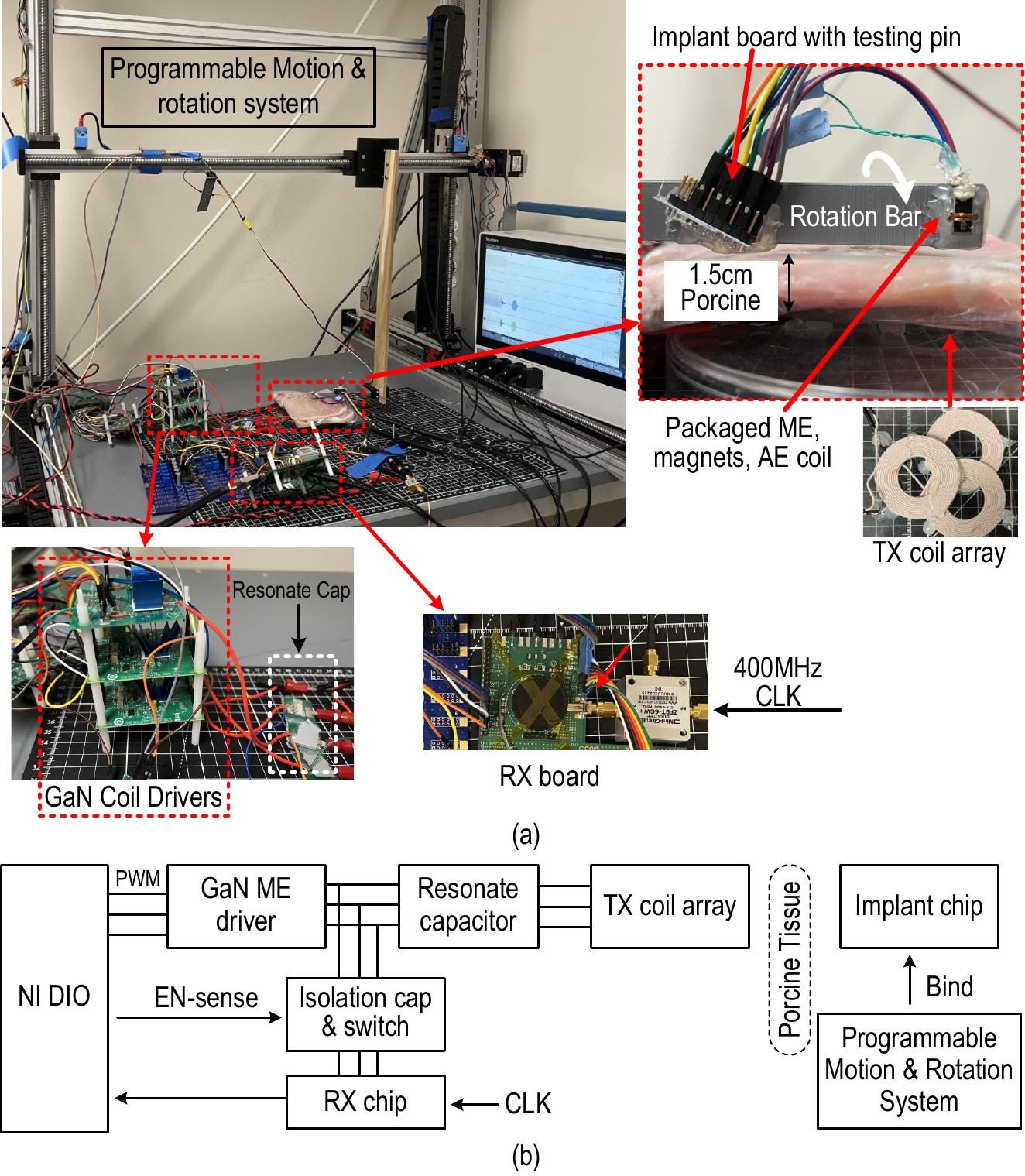}%
}
\caption{(a) Testing setup; (b) system topology under testing.}
\label{testing_setup}
\end{figure}

We performed comprehensive in-vitro experiments with the complete system enabled by the two chips. Fig.~\ref{testing_setup} displays our testing setup and physical connections. Three off-the-shelf GaN half-bridge drivers (MASTERGAN1 from STM) were used to drive the TX LC tanks efficiently. The half-bridge configuration ensures a grounded connection during the AE sensing phase. The implant chip packaged with an ME film and an AE coil was installed on a motor stage capable of moving in three directions and rotating in an X-Z plane via a servo motor. A 3D-printed plastic rotation bar extends from the rotation axis of the servo motor to mount the implant under test. 
A wood bar connector linked the implant's rotation motor to the connecting point of the motion system, minimizing interference from the motion system. 
A 1.5-cm porcine tissue was placed between three TX coils and the implant board. On the receiver side, the RX chip was connected to the output of the ME driver through an isolation capacitor. This isolation capacitor creates a high pass corner with the input impedance of the AE RX to remove the DC draft of the GaN driver's output. Additionally,  a protection switch was inserted at the beginning of the receiver chain and controlled by NI DIO, synchronizing with the TX driver. The DIO collected the three digitized amplitudes and three polarities from the RX chip and provided the desired PWM control signal to the drivers.

\begin{figure}[t!]
\centering
\includegraphics[width=\linewidth]{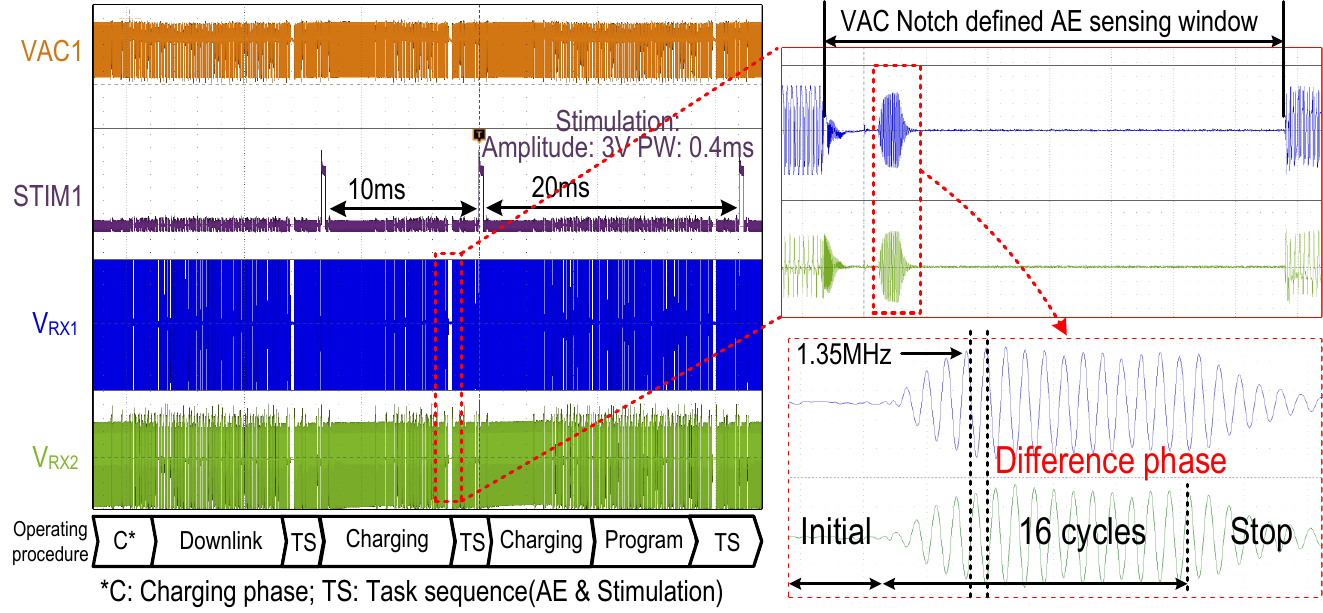}
\vskip -1ex
\caption{System operation waveform when the implant is under 90$^\circ$ rotation.}
\label{working_waveform}
\end{figure}

Fig.~\ref{working_waveform} shows the measured operation waveforms over an entire operation cycle, when the implant is rotated by 90\textdegree{}. As described in Fig.~\ref{system_overview}~b, the system cycles through charging, downlink programming, AE sensing, and stimulation. 
{Here the bi-phasic stimulator is programmed to 3V 0.4ms pulse width with 10ms and 20ms intervals. }
The zoom-in view in Fig.~\ref{working_waveform} shows the two AE RX channels capturing the 1.35MHz AE signal with opposite polarities. 

\begin{figure}[t!]
\centering
\includegraphics[width=\linewidth]{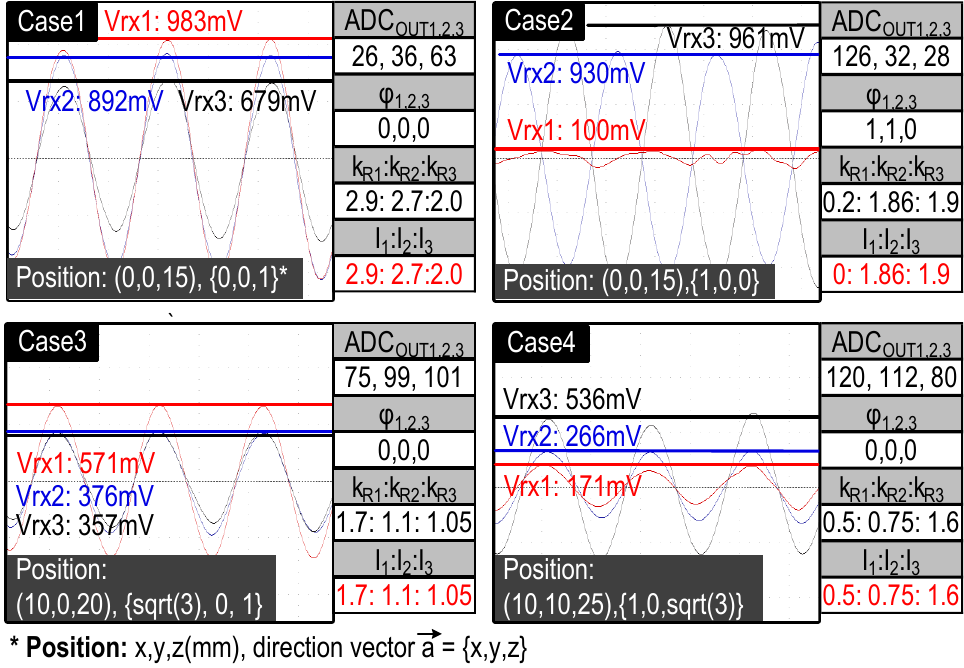}
\vskip -1ex
\caption{Measured 3-channel RX outputs (coupling coefficients, phases, and TX coil array configuration) at various positions.}
\label{four_case_rotation}
\end{figure}

We evaluated the closed-loop WPT performance across various implant positions and orientations in \textit{in-vitro} condition, shown in Fig.~\ref{four_case_rotation}. In different cases, the amplitude and polarity detection were working properly, and the smaller ADC output value indicated a larger amplitude, which could be concluded first as the reference clock of polarity detection. 
When the coupling coefficient of one channel is significantly lower than another, falling below the empirical threshold of 8, we will deactivate that driver. This is because PWM control lacks precision under these conditions, and the static power consumption of the channel is relatively high, while its contribution to magnetic field generation is minor. 

\begin{figure}[!t]
\centering
\includegraphics[width=0.8\linewidth]{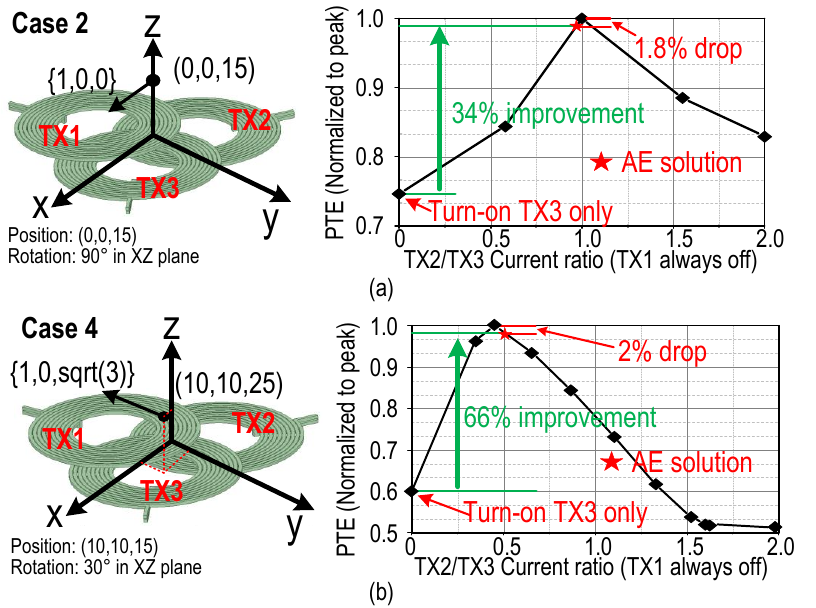}
\vskip -2ex
\caption{Normalized PTE from peak point when current in TX2 and TX3 sweep under (a) Case 2 when the implant is placed at 15mm above the origin and rotated in XZ-plane by 90\textdegree{} from the ideal orientation, and (b) case 4 when the implant has an arbitrary set of lateral and angular misalignment (position: 10mm,10mm,25mm; rotation vector: \{1, 0, $\sqrt{3}$\})}
\label{Case_PTE_manual_sweep}
\vskip -3ex
\end{figure}

In order to confirm that the AE solution is indeed optimal, we manually sweep the current ratio in TX coils in two representative cases (2 and 4). For simplicity, we turn off the TX1 in this test and normalize the PTE to the peak point in each case, as shown in Fig.~\ref{Case_PTE_manual_sweep}. In case 2, the special arrangement allows for a simple theoretical calculation of the optimal ratio between TX2 and TX3, which is 1. The manual sweeping test agrees with the calculation. The AE solution is merely 1.8\% away from the PTE achieved by ideal control.
In case 4, the implant has an arbitrary set of lateral and angular misalignments. Guided by theoretical analysis, we took finer sweeping steps to find the optimal condition in the testing. Again, the AE scheme achieves the expected result and significantly increases the efficiency of power transfer by 66\% with only a 2\% drop from the highest PTE found by manual sweeping. This tracking error is mainly incurred by the non-linearity of the AE RX chain, because we use an open-loop structure for the front end to reduce noise and power. 

\begin{figure}[!t]
\centering
\includegraphics[width=\linewidth]{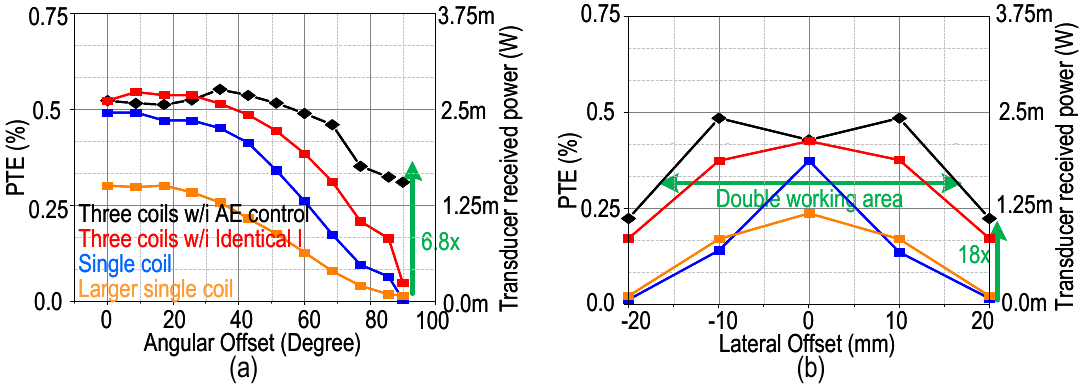}
\vskip -2ex
\caption{
{Measured power transfer efficiency under (a) angular rotation test, where the implant is located at (0,0,20) and rotates from \{0,0,1\} to \{1,0,0\}; (b) lateral offset test, where the implant moves from (-20,10,20) to (20,10,20) and keeps the direction of \{0,0,1\}.}}
\label{efficiency}
\vskip -3ex
\end{figure}

The PTE measurements in Fig.~\ref{efficiency} demonstrate how AE compares with three baselines in terms of sensitivity to rotational and lateral misalignment. 
{The baselines include 1) a single coil in our optimized 3-coil array placed at the origin (diameter = 4.2cm); 2) a single coil with roughly the same footprint of three coils combined (diameter = 6.8cm); and 3) the same three coils as AE but with constant non-adaptive driving currents. We reuse coils for these baselines with the intention of reducing the impacts of coil designs (parameters, number of turns, etc.) on field distribution and PTE measurements.}
It can be observed that three coils, even without adaptation, outperform the single-coil baseline with consistently higher PTE over both lateral and rotational mismatch. AE offers superior PTE across all tested poses, the improvements are more significant under severe misalignment. Compared to the single-coil baseline, AE improves the PTE by 6.8$\times$ with 90\textdegree{} rotation from the ideal orientation and by 18$\times$ with 20mm lateral offset. 


\subsection{Comparison with Other Implantable WPT Systems}

\begin{table*}[t]
\caption{\textbf{Comparisons with State-of-the-Art Wireless Power Transfer Technologies for Miniature Biomedical Implants}}
\label{table}
\centering
\setlength{\tabcolsep}{2pt}
\renewcommand{\arraystretch}{1.5}
\begin{tabular}{|p{70pt}|p{50pt}|p{55pt}|p{55pt}|p{55pt}|p{55pt}|p{60pt}|p{60pt}|}
\hline

& 
\parbox[c][0.8cm]{50pt}{\centering
{\textbf{This Work}}}&
\parbox[c][0.8cm]{55pt} {\centering{Z. Yu \\RFIC'22\cite{yu_wireless_2022}}}&
\parbox[c][0.8cm]{55pt} {\centering{Z. Yu\\JSSC'22\cite{yu_magnetoelectric_2022}}}&
\parbox[c][0.8cm]{55pt} {\centering{J. Lee\\Nat. Elec'21\cite{lee_neural_2021}}} &
\parbox[c][0.8cm]{55pt} {\centering{I.Habibagahi\\Sci. Rep.'22\cite{habibagahi_vagus_2022}}} 
&
\parbox[c][0.8cm]{60pt}{\centering{D. Piech\\Nat. BME'20\cite{piech_wireless_2020}}} &
\parbox[c][0.8cm]{60pt}{\centering{J. Charthad\\TBioCAS'18\cite{charthad_mm-sized_2018}}} 
\\\hline

\parbox[c][0.8cm]{70pt}{\centering
{\textbf{WPT Mechanism}}} &
\parbox[c][0.8cm]{50pt}{\centering
{\textbf{ME\\(340kHz)}}}&
\parbox[c][0.8cm]{55pt} {\centering{ME\\(340kHz)}}&
\parbox[c][0.8cm]{55pt} {\centering{ME\\(330kHz)}}&
\parbox[c][0.8cm]{55pt}{\centering
{Inductive\\(900MHz)}}&
\parbox[c][0.8cm]{55pt} {\centering{Inductive\\(13.56MHz)}}  &
\parbox[c][0.8cm]{60pt} {\centering{Ultrasound\\(1.85MHz)}}  &
\parbox[c][0.8cm]{60pt} {\centering{Ultrasound\\(1.314MHz)}} 
\\\hline

\parbox[c][0.8cm]{70pt}{\centering
{\textbf{TX Type}}} &
\parbox[c][0.8cm]{50pt}{\centering
{\textbf{Planar Coil Array}}}&
\parbox[c][0.8cm]{55pt} {\centering{TX Coil + Sense Coil}}&
\parbox[c][0.8cm]{55pt} {\centering{Single Coil}}&
\parbox[c][0.8cm]{55pt} {\centering{TX Coil + Relay Coil}}  &
\parbox[c][0.8cm]{55pt} {\centering{Single Coil}}  &
\parbox[c][0.8cm]{60pt}{\centering
{Single Transducer}}&
\parbox[c][0.8cm]{60pt} {\centering{Single Transducer}} 
\\\hline

\parbox[c][0.8cm]{70pt}{\centering
{\textbf{Power Transducer\\Dimension (mm)}}} &
\parbox[c][0.8cm]{50pt}{\centering
{\textbf{5x2x0.18}}}&
\parbox[c][0.8cm]{55pt} {\centering{4x2x0.1}}&
\parbox[c][0.8cm]{55pt} {\centering{4x2x0.1}}&
\parbox[c][0.8cm]{55pt} {\centering{0.5x0.5\\(On-chip)}}  &
\parbox[c][0.8cm]{55pt} {\centering{15x15x0.025\\(Flexible PCB)}}  &
\parbox[c][0.8cm]{60pt}{\centering
{1x0.8x0.8}}&
\parbox[c][0.8cm]{60pt} {\centering{1.65x1.5x1.5}} 
\\\hline

\parbox[c][0.8cm]{70pt}{\centering
{\textbf{Global Regulation}}} &
\parbox[c][0.8cm]{50pt}{\centering
{\textbf{Yes\\Active Echo}}}&
\parbox[c][0.8cm]{55pt} {\centering{Yes\\LSK Backscatter}}&
\parbox[c][0.8cm]{55pt} {\centering{No}}&
\parbox[c][0.8cm]{55pt} {\centering{No}}  &
\parbox[c][0.8cm]{55pt} {\centering{No}} &
\parbox[c][0.8cm]{60pt}{\centering
{No}}&
\parbox[c][0.8cm]{60pt} {\centering{No}} 
\\\hline

\multicolumn{1}{|c|}{\textbf{Technology (nm)}} 
& \multicolumn{1}{c|}{\textbf{180}} 
& \multicolumn{1}{c|}{180} 
& \multicolumn{1}{c|}{180}
& \multicolumn{1}{c|}{65}
& \multicolumn{1}{c|}{180}
& \multicolumn{1}{c|}{65}
& \multicolumn{1}{c|}{180HV}
\\\hline

\multicolumn{1}{|c|}{\textbf{Bio-Application}}  
& \multicolumn{1}{c|}{\textbf{Stimulation}} 
& \multicolumn{1}{c|}{Stimulation} 
& \multicolumn{1}{c|}{Stimulation}
& \multicolumn{1}{c|}{Record./Stim.}
& \multicolumn{1}{c|}{Stimulation}
& \multicolumn{1}{c|}{Stimulation}
& \multicolumn{1}{c|}{Stimulation}
\\\hline

\parbox[c][0.8cm]{70pt}{\centering
{\textbf{Omnidirectional}}} &
\parbox[c][0.8cm]{50pt}{\centering
{\textbf{Yes}}}&
\parbox[c][0.8cm]{55pt} {\centering{No}}&
\parbox[c][0.8cm]{55pt} {\centering{No\\(\textless 50$^\circ$ rotation)}}&
\parbox[c][0.8cm]{55pt} {\centering{No}}  &
\parbox[c][0.8cm]{55pt} {\centering{No}} &
\parbox[c][0.8cm]{60pt}{\centering
{No\\(\textless45$^\circ$ rotation)}}&
\parbox[c][0.8cm]{60pt} {\centering{No\\\textless45$^\circ$ rotation)}} 
\\\hline

\parbox[c][0.8cm]{70pt}{\centering
{\textbf{Peak PTE$^\#$\\(Aligned)$^{*}$}}} &
\parbox[c][0.8cm]{50pt}{\centering
{\textbf{0.55\%\\(20mm)}}}&
\parbox[c][0.8cm]{55pt} {\centering{0.2\%\\(20mm)}}&
\parbox[c][0.8cm]{55pt} {\centering{0.43\%\\(20mm)}}&
\parbox[c][0.8cm]{55pt} {\centering{0.08\%\\(8mm)}}  &
\parbox[c][0.8cm]{55pt} {\centering{18\%$^\$$\\(30mm)}} &
\parbox[c][0.8cm]{60pt}{\centering
{0.073\%\\(18mm)}}&
\parbox[c][0.8cm]{60pt} {\centering{0.04\%\\(105mm)}} 
\\\hline

\parbox[c][0.8cm]{70pt}{\centering
{\textbf{Peak PTE$^\#$\\(90$^\circ$ Rotation)$^{*}$}}} &
\parbox[c][0.8cm]{50pt}{\centering
{\textbf{0.31\%\\(20mm)}}}&
\parbox[c][0.8cm]{55pt} {\centering{N/A}}&
\parbox[c][0.8cm]{55pt} {\centering{0.045\%\\(20mm)$^\$$}}&
\parbox[c][0.8cm]{55pt} {\centering{N/A}}  &
\parbox[c][0.8cm]{55pt} {\centering{0$^\$$}} &
\parbox[c][0.8cm]{60pt} {\centering{N/A}}&
\parbox[c][0.8cm]{60pt} {\centering{N/A}} 
\\\hline

\parbox[c][0.8cm]{70pt}{\centering
{\textbf{Maximum Received Power$^{*}$}}} &
\parbox[c][0.8cm]{50pt}{\centering
{\textbf{2.75mW\\(20mm)}}}&
\parbox[c][0.8cm]{55pt} {\centering{N/A}}&
\parbox[c][0.8cm]{55pt} {\centering{N/A}}&
\parbox[c][0.8cm]{55pt} {\centering{0.025mW\\(8mm)}}  &
\parbox[c][0.8cm]{55pt} {\centering{N/A}} &
\parbox[c][0.8cm]{60pt} {\centering{0.0118mW\\(18mm)}}&
\parbox[c][0.8cm]{60pt} {\centering{5.94mW\\(10.5mm)}} 
\\\hline



\end{tabular}

\begin{tablenotes}
\footnotesize
\item[]
{$^\#$ PTE is defined as the maximum AC power received by RX transducer divided by the AC power consumption of the TX transducer;} \\
$^{*}$ TX-RX distance reported in (); $^\$$ Read from figure.

\end{tablenotes}
\label{tab1}
\end{table*}
Table \ref{tab1} compares our system with prior implantable systems using ME\cite{yu_wireless_2022, yu_magnetoelectric_2022}, inductive\cite{lee_neural_2021, tang_336_2021}, and ultrasonic transducers\cite{piech_wireless_2020,charthad_mm-sized_2018} for wireless power transfer. We achieve omnidirectional power based on active echo technology, sustaining operation at all rotation degrees, and the PTE only drops less than 50\% compared with ideal alignment, which is not supported in the prior art designs.

%
\section{Conclusion}

In conclusion, this paper presents the Active Echo (AE) technology to enable omnidirectional wireless power transfer (WPT) for millimetric magnetoelectric bioelectronic implants. 
We analytically and experimentally prove the feasibility of using an auxiliary coil on the implant to adaptively control a multi-coil TX array for optimal magnetic flux steering, overcoming the primary challenges of coupling sensing and adaptive multi-coil WPT with weakly coupled TX and RX. 
We demonstrated a complete bioelectronic implant system, including a 14.2mm$^3$ implantable neurostimulator with a custom SoC powered by a magnetoelectric (ME) film, an external transmitter including a custom AE receiver chip, and a mutual inductance canceled multi-coil TX array. We built two custom chips with various circuit techniques to efficiently and reliably perform AE signal transmission and reception. 
The prototyping system functions properly in a fully wireless in-vitro testing setup and demonstrates omnidirectional power transfer capabilities, showcasing a 6.8x higher PTE than a single-coil baseline under 90\textdegree{} rotation from the ideal alignment, as well as less than 2\% efficiency drop compared to using ideal control.





\section*{Acknowledgment}
The authors would like to thank Huan-Cheng Liao, Zhiyu Chen, Fatima T. Alrashdan, Yan He, and Xi Hu for valuable discussions and support.

\ifCLASSOPTIONcaptionsoff
  \newpage
\fi

\bibliography{bib/JSSC_multi_coil.bib,bib/Revision1.bib, bib/bstcontrol}
\bibliographystyle{IEEEtran}

\begin{IEEEbiography}[{\includegraphics[width=1in,height=1.25in,clip,keepaspectratio]{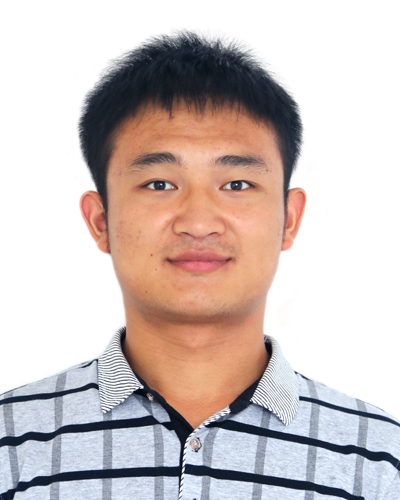}}]{Wei Wang}
(Graduate Student Member, IEEE) received a B.S. degree in electronic information science and technology from the Harbin Institute of Technology, Harbin, China, in 2016, and an M.S. degree in integrated circuit engineering from Tsinghua University, Beijing, China, in 2019. He is currently pursuing the Ph.D. degree in electrical
and computer engineering with Rice University, Houston, TX, USA.
His research interests include mixed-signal circuits and systems design.
\end{IEEEbiography}

\vskip -1\baselineskip
\begin{IEEEbiography}[{\includegraphics[width=1in,height=1.25in,clip,keepaspectratio]{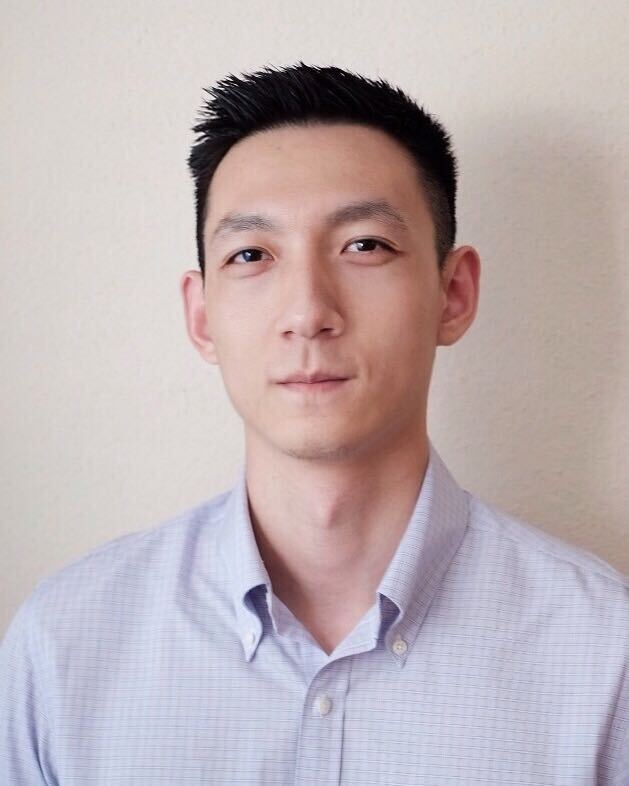}}]{Zhanghao Yu} received a B.E. degree in Integrated Circuit Design and Integrated Systems from the University of Electronic Science and Technology of China, Chengdu, China, in 2016, and an M.S. degree in Electrical Engineering from the University of Southern California, Los Angeles, CA, USA, in 2018. He earned his Ph.D. degree in Electrical and Computer Engineering from Rice University, Houston, TX, USA, in 2023. He worked as an analog design engineering intern at Analog Devices Inc. in 2022 and has been with Intel as an Analog Design Engineer since 2023. His research interests include analog and mixed-signal circuit design for wireless implantable bioelectronics, clocking, power management, wireless power transfer, and low-power communication. Dr. Yu was a recipient of the 2024 SSCS Rising Stars Award and 2021-2022 SSCS Predoctoral Achievement Award. He received Best Paper Awards at 2021 CICC and 2022 MobiCom, and the Best Student Paper Finalist at 2022 RFIC.
\end{IEEEbiography}

\vskip -1\baselineskip
\begin{IEEEbiography}[{\includegraphics[width=1in,height=1.25in,clip,keepaspectratio]{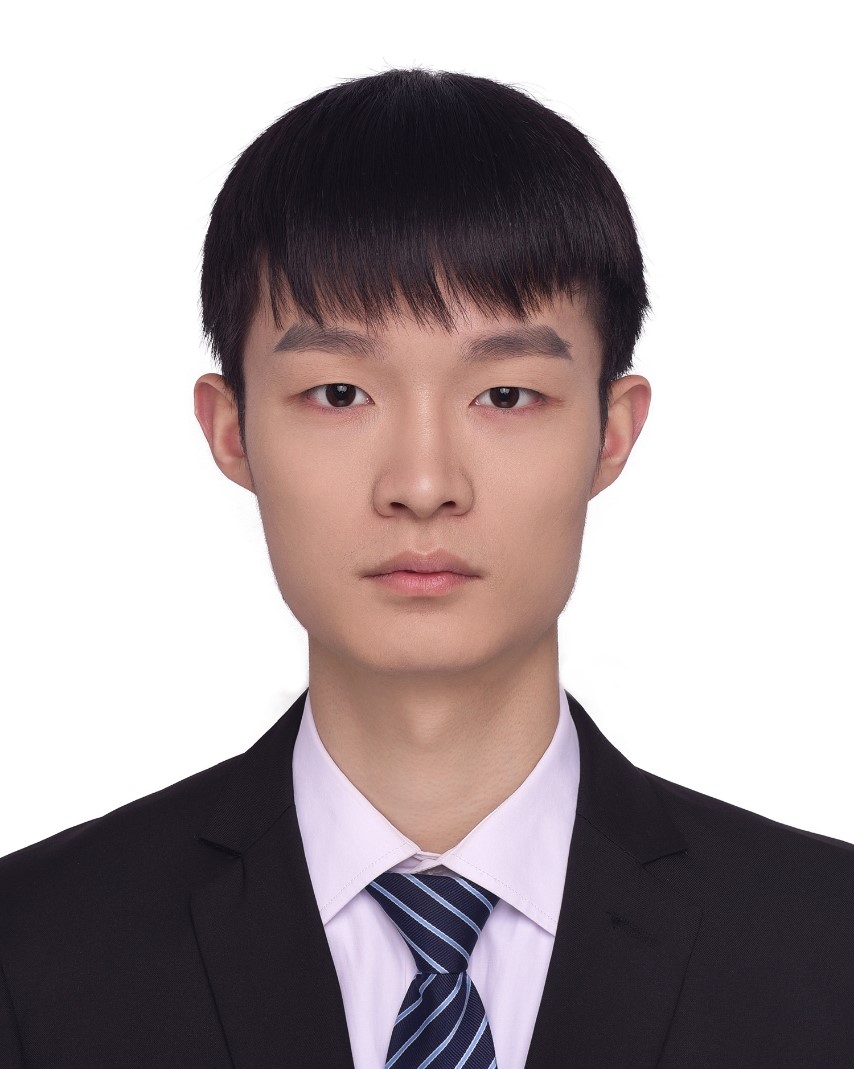}}]{Yiwei Zou} (Graduate Student Member, IEEE) received the B.E. degree in Integrated Circuits and Systems from Huazhong University of Science and Technology, Wuhan, China, in 2022. He is currently working toward his Ph.D. degree in Electrical and Computer Engineering at Rice University, Houston, TX. His research interests include analog and mixed-signal integrated circuits design for power management and bio-electronics.
\end{IEEEbiography}

\vskip -1\baselineskip
\begin{IEEEbiography}
[{\includegraphics[width=1in,height=1.25in,clip,keepaspectratio]{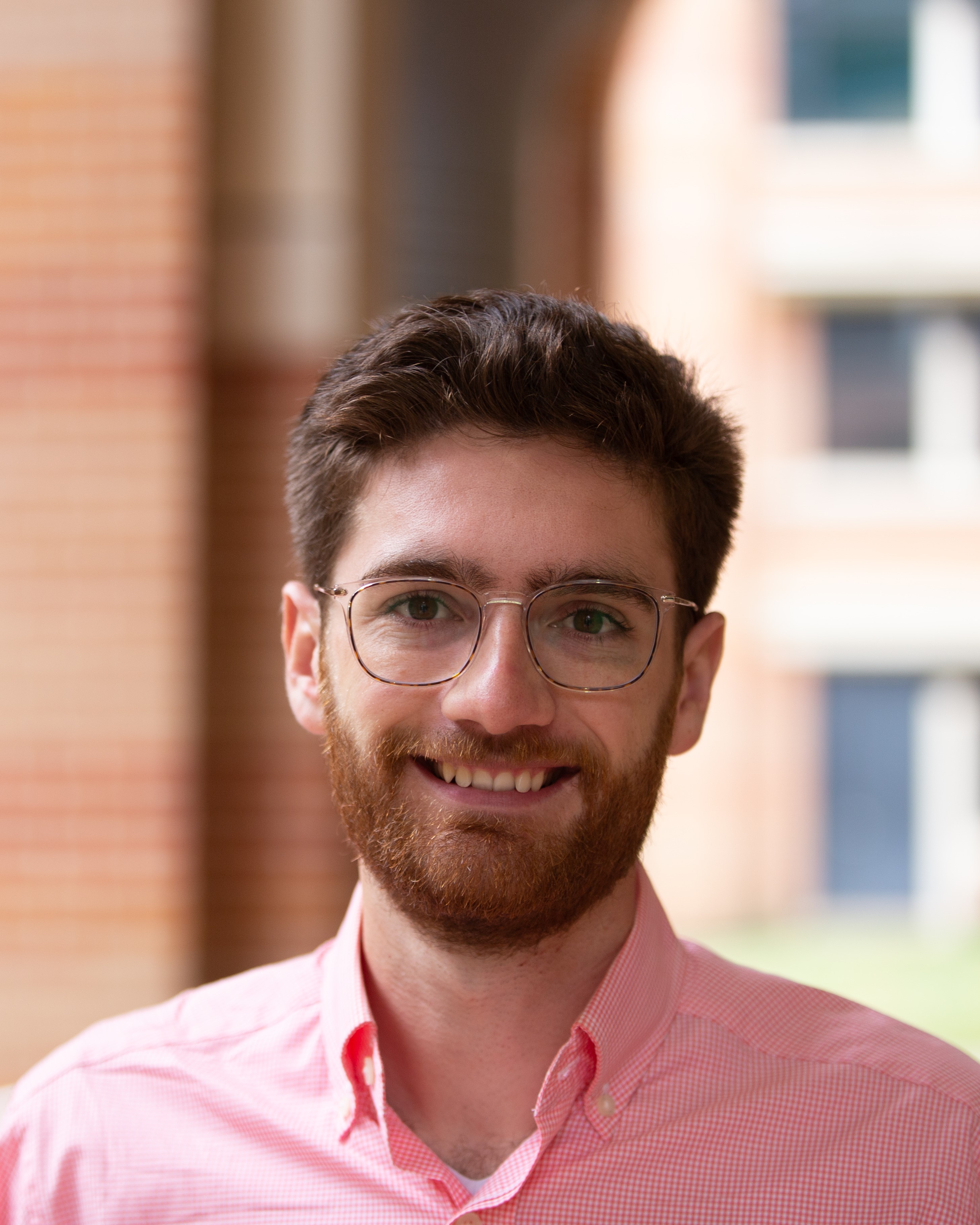}}]{Joshua E. Woods} received a B.S. in Electrical Engineering from the University of Michigan Ann Arbor in 2021. He is currently pursuing a PhD in Electrical Engineering at Rice University in Houston, TX. His research interests include wireless power, embedded systems, and closed loop bioelectronic systems. 
\end{IEEEbiography}

\begin{IEEEbiography}
[{\includegraphics[width=1in,height=1.25in,clip,keepaspectratio]{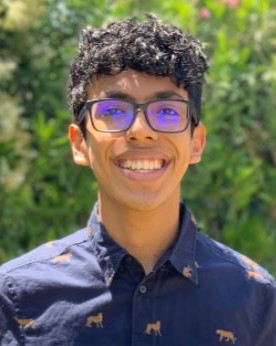}}]{Prahalad Chari} (Undergraduate Student Member, IEEE) is currently pursuing the BS degree in electrical and computer engineering from Rice University, Houston, TX, USA, and will graduate in 2025. He is looking to pursue a Ph.D degree in computer engineering. His research interests include digital circuit design and computer architecture.
\end{IEEEbiography}

\vskip -1\baselineskip
\begin{IEEEbiography}
[{\includegraphics[width=1in,height=1.25in,clip,keepaspectratio]{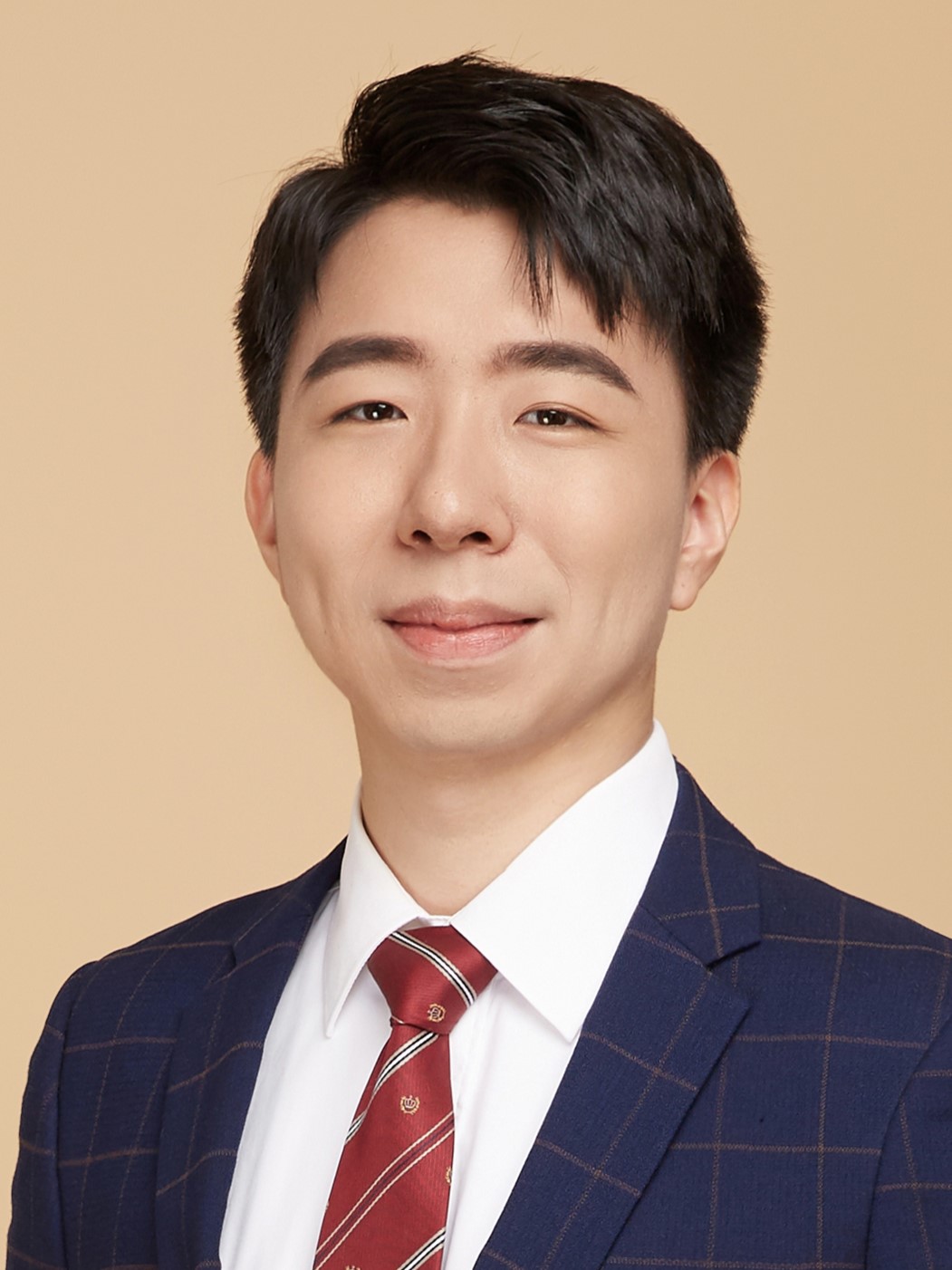}}]{Yumin Su} (Student Member, IEEE) received the B.S. degree in electrical and computer engineering from Rice University, Houston, TX, USA, in 2023. He is currently pursuing a Ph.D. degree in electrical and computer engineering at Rice University, Houston, TX, USA. 
His research interests include hardware security and design automation.
\end{IEEEbiography}

\vskip -1\baselineskip
\begin{IEEEbiography}[{\includegraphics[width=1in,height=1.25in,clip,keepaspectratio]{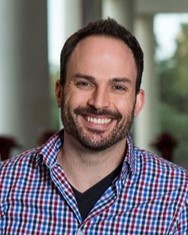}}]{Jacob Robinson} (Senior Member, IEEE) is a Professor in Electrical \& Computer Engineering and Bioengineering at Rice University, where his group develops miniature technologies to manipulate and monitor neural circuit activity. He received a B.S. in Physics from UCLA, a Ph.D. in Applied Physics from Cornell University, and completed Postdoctoral training in the Chemistry Department at Harvard. He previously served as the co-chair of the IEEE Brain Initiative and a core member of the IEEE Brain Neuroethics working group and is currently a member of the IEEE EMBS AdCom. In addition to his academic work, Dr. Robinson is the co-founder and CEO of Motif Neurotech, which is developing minimally invasive bioelectronics to treat mental health disorders.
\end{IEEEbiography}

\vskip -1\baselineskip
\begin{IEEEbiography}
[{\includegraphics[width=1in,height=1.25in,clip,keepaspectratio]{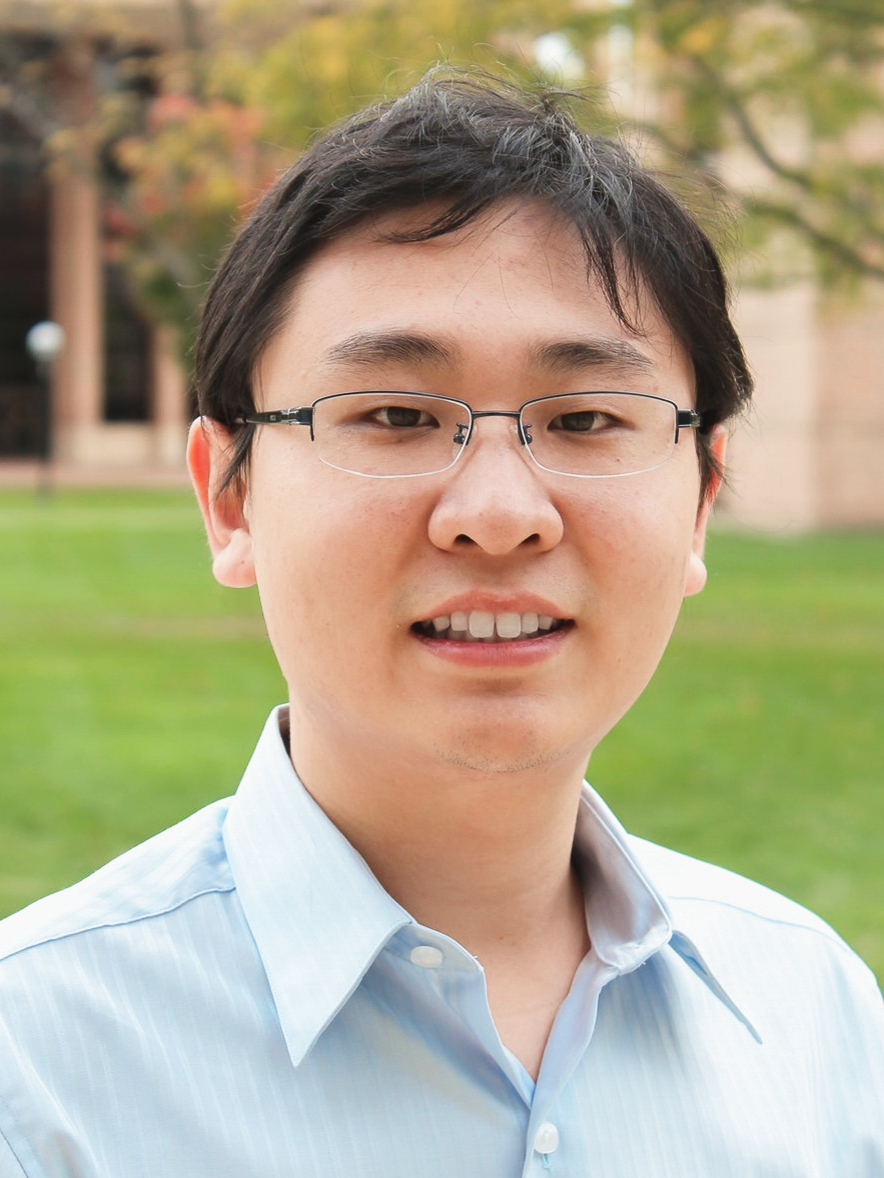}}] {Kaiyuan Yang} (Member, IEEE) is an Associate Professor of Electrical and Computer Engineering at Rice University, USA, where he leads the Secure and Intelligent Micro-Systems (SIMS) lab. He received a B.S. degree in Electronic Engineering from Tsinghua University, China, in 2012, and a Ph.D. degree in Electrical Engineering from the University of Michigan - Ann Arbor, MI, in 2017. His research focuses on low-power integrated circuits and system design for bioelectronics, hardware security, and mixed-signal/in-memory computing. 

Dr. Yang is a recipient of National Science Foundation CAREER Award, IEEE SSCS Predoctoral Achievement Award, and best paper awards from premier conferences in multiple fields, including 2024 Annual International Conference of the IEEE Engineering in Medicine and Biology Society (EMBC), 2022 ACM Annual International Conference on Mobile Computing and Networking (MobiCom), 2021 IEEE Custom Integrated Circuit Conference (CICC), 2016 IEEE International Symposium on Security and Privacy (Oakland), and 2015 IEEE International Symposium on Circuits and Systems (ISCAS). His research was also selected as the research highlights of Communications of ACM and ACM GetMobile magazines, and IEEE Top Picks in Hardware and Embedded Security. He is currently serving as an associate editor of IEEE Transactions on VLSI Systems (TVLSI) and a program committee member of ISSCC, CICC, and DAC conferences. 
\end{IEEEbiography}

\vspace{11pt}

\vspace{11pt}

\vfill

\end{document}